\documentclass[toc]{PoS}

\usepackage{amssymb}
\usepackage{amsmath}
\usepackage{amsfonts}
\usepackage{bbm}
\usepackage{amsthm}
\usepackage{mathrsfs}
\usepackage{color}

\usepackage{mathrsfs}
\usepackage{bm}
\usepackage[arrow,matrix,curve]{xy}

\usepackage{amstext,amscd,array}

\theoremstyle{plain}

\theoremstyle{definition}

\numberwithin{equation}{section}

\def\dd{\mathrm{d}}

\def\bbH{\mathbb{H}}

\newcommand{\obultimes}{\mathbin{\ooalign{$\otimes$\cr\hidewidth\raise0.17ex\hbox{$\scriptstyle\bullet\mkern4.48mu$}}}}
\newcommand{\ostartimes}{\mathbin{\ooalign{$\otimes$\cr\hidewidth\raise0.17ex\hbox{$\scriptstyle\star\mkern4.48mu$}}}}

\newcommand{\obulplus}{\mathbin{\ooalign{$\boxplus$\cr\hidewidth\raise0.295ex\hbox{$\scriptstyle\bullet\mkern4.7mu$}}}}

\usepackage{epsf}
\usepackage{epsfig}
\usepackage{wrapfig}

\input{epsf}

\def\tri{\triangleright}

\newcommand{\mbf}[1]{{\boldsymbol {#1} }}
\def\ii{{\,{\rm i}\,}}
\def\dd{{\rm d}}

\def\A{{\sf A}}
\def\V{{\sf V}}

\def\tri{{\,{\scriptstyle\triangle}\,}}

\def\mfg{{\mathfrak g}}

\def\mcH{{\mathcal H}}

\newcommand{\CCP}{\mathscr{P}}

\newcommand{\CX}{\mathcal{X}}

\newcommand{\CJ}{\mathcal{J}}
\newcommand{\CU}{\mathcal{U}}

\newcommand{\eq}{\begin{equation}}
\newcommand{\eqend}{\end{equation}}
\newcommand{\eqa}{\begin{eqnarray}}
\newcommand{\nonueqa}{\begin{eqnarray*}}
\newcommand{\eqaend}{\end{eqnarray}}
\newcommand{\nonueqaend}{\end{eqnarray*}}

\newcommand{\bma}[1]{\begin{array}{#1}}
\newcommand{\ema}{\end{array}}
\newcommand{\bc}{\begin{center}}
\newcommand{\ec}{\end{center}}

\newcommand{\complex}{{\mathbb C}} %% complex numbers
\newcommand{\zed}{{\mathbb Z}} %% integers
\newcommand{\nat}{{\mathbb N}} %% naturals
\newcommand{\real}{{\mathbb R}} %% real numbers

\def\Tcal{{\mathcal T}}

\newif\ifold             \oldtrue

\def\e{{\,\rm e}\,}

\def\be{\begin{equation}}
\def\ee{\end{equation}}
\def\bea{\begin{eqnarray}}
\def\eea{\end{eqnarray}}
\def\bd{\begin{displaymath}}
\def\ed{\end{displaymath}}

\newcommand{\beq}{\begin{eqnarray}}
\newcommand{\eeq}{\end{eqnarray}}

\newcommand{\co}{{\mathcal O}}

\makeatletter
\newdimen\normalarrayskip              % skip between lines
\newdimen\minarrayskip                 % minimal skip between lines
\normalarrayskip\baselineskip
\minarrayskip\jot
\newif\ifold             \oldtrue            
\def\arraymode{\ifold\relax\else\displaystyle\fi} % mode of array entries
\def\@arrayskip{\ifold\baselineskip\z@\lineskip\z@
     \else
     \baselineskip\minarrayskip\lineskip2\minarrayskip\fi}
\def\@arrayclassz{\ifcase \@lastchclass \@acolampacol \or
\@ampacol \or \or \or \@addamp \or
   \@acolampacol \or \@firstampfalse \@acol \fi
\edef\@preamble{\@preamble
  \ifcase \@chnum
     \hfil$\relax\arraymode\@sharp$\hfil
     \or $\relax\arraymode\@sharp$\hfil
     \or \hfil$\relax\arraymode\@sharp$\fi}}
\def\@array[#1]#2{\setbox\@arstrutbox=\hbox{\vrule
     height\arraystretch \ht\strutbox
     depth\arraystretch \dp\strutbox
     width\z@}\@mkpream{#2}\edef\@preamble{\halign \noexpand\@halignto
\bgroup \tabskip\z@ \@arstrut \@preamble \tabskip\z@ \cr}%
\let\@startpbox\@@startpbox \let\@endpbox\@@endpbox
  \if #1t\vtop \else \if#1b\vbox \else \vcenter \fi\fi
  \bgroup \let\par\relax
  \let\@sharp##\let\protect\relax
  \@arrayskip\@preamble}
\makeatother

\def\be{\beta}

\theoremstyle{definition}

\title{Higher Quantum Geometry and \\ Non-Geometric String Theory}

\ShortTitle{Higher Quantum Geometry and Non-Geometric String Theory}

\author{\speaker{Richard J. Szabo}%
        \vspace{2mm}\\
Department of Mathematics, Heriot-Watt University, Edinburgh, United Kingdom.\vspace{1mm}\\
Maxwell Institute for Mathematical Sciences, Edinburgh, United Kingdom.\vspace{1mm}\\
The Higgs Centre for Theoretical Physics, Edinburgh, United Kingdom.\vspace{2mm}\\
E-mail: \email{R.J.Szabo@hw.ac.uk}}

\abstract{We present a concise overview of the physical and mathematical structures underpinning the appearence of nonassociative deformations of geometry in non-geometric string theory. Starting from a quick recap of the appearence of noncommutative product and commutator deformations of geometry in open string theory with $B$-fields, we argue on physical principles that closed strings should instead probe triproduct and tribracket deformations in backgrounds of locally non-geometric fluxes. After describing the toy model of electric charges moving in fields of smooth distributions of magnetic charge as a physical introduction to the notions of nonassociative geometry, we review the description of non-geometric fluxes in generalized geometry and double field theory, and the worldsheet calculations suggesting the appearence of nonassociative deformations, together with their caveats. We discuss how algebroids and their associated AKSZ sigma-models give a description of non-geometric backgrounds in terms of higher geometry, and consider the quantization of the membrane sigma-model which geometrizes closed strings with $R$-flux. From this we derive an explicit nonassociative star product for the quantum geometry of the closed string phase space, and apply it to derive the triproducts that appear in conformal field theory correlation functions, to describe a consistent treatment of nonassociative quantum mechanics, to demonstrate quantitatively the coarse-graining of spacetime due to $R$-flux, and to describe the quantization of Nambu brackets. We also briefly review how these constructions lead to a nonassociative theory of gravity, their uplifts to non-geometric M-theory, and the role played by $L_\infty$-algebras in these developments.}

\FullConference{Proceedings of the Corfu Summer Institute 2017 ``School and Workshops on
Elementary Particle Physics and Gravity''\\
                 2--28 September 2017\\
                 Corfu, Greece\\
{\small Preprint: \  {\tt EMPG--18--08}}}

\begin{document}

%%%%%%%%%%%%%%%%%%%%%%%%%%%%%%%%%%%%%%%%%%%%%%%%%%%%%%%
%%%%%%%%%%%%%%%%%%%%%%%%%%%%%%%%%%%%%%%%%%%%%%%%%%%%%%%

\section{Introduction: String theory and higher noncommutative geometry}
\label{sec:intro}

The second Training School of the COST Action {\sl Quantum Structure
  of Spacetime} was devoted to the general topic ``Quantum Spacetime
and Physics Models''. While this is a broad topic with many potential directions (some covered by other lectures at this School), 
for the purposes of these lectures this refers to the
problem that Einstein's general theory of relativity cannot be
consistently quantized via quantum field theory, due to the
ultraviolet divergences that plague perturbation theory around a flat
background which require infinitely-many counterterms. To attempt to
solve this problem one can consider physics models with a natural minimal
length providing a suitable ultraviolet regularization. In this
contribution we consider two such theories and how to reconcile them
within the context of the School. Our treatment in this section and some other portions of this article
have been influenced by various other reviews available, such as e.g.~\cite{Lust:2012fp,Plauschinn:2012kd,Mylonasa,Blumenhagen:2014sba}. 

One approach is based on noncommutative geometry. In attempting to
reconcile quantum mechanics with gravity, which is a theory based on
the geometry of spacetime, one is inevitably led to the notion of
`quantum geometry', which refers to the application of the principles
of quantum mechanics to spacetime itself. One way to think of such a
quantization is by promoting the spacetime coordinates $x^i$ to
`operators' which do not commute:
\bea\label{eq:xixjthetaij}
[x^i,x^j]=\ii\hbar\,\theta^{ij} \ ,
\eea
for some bivector $\theta^{ij}$ which naturally
incorporates a minimal area: Applying the standard uncertainty
principle to the commutation relations \eqref{eq:xixjthetaij} implies $\Delta x^i\, \Delta
x^j\geqslant\frac\hbar2\, |\theta^{ij}|$, so that the minimal value of
$|\theta^{ij}|$ may be thought of as the Planck length $\ell_{\rm P}$
of spacetime. Quantum field theory on such noncommutative spaces
exhibits very interesting features of forbidden interactions,
controlled Lorentz violation, and UV/IR mixing (see
e.g.~\cite{Szabosignatures} for a review), and it can be extended to a
noncommutative theory of gravity~\cite{Aschieri:2005yw,AschieriNCgrav} as discussed in the lectures of L.~Castellani at this School. However, there are two major pitfalls to this
approach. Firstly, although a minimal length scale is naturally
introduced, there is no coarse-graining of spacetime that one would
expect from a quantum theory of gravity: An underlying discrete
structure such as a quantum of minimal volume does not appear in this
framework. Secondly, most treatments assume that the brackets
\eqref{eq:xixjthetaij} satisfy the Jacobi identity; in particular,
when the bivector $\theta=B^{-1}$ is invertible, it is not clear how to deal with
the cases with non-vanishing flux $H=\dd B\neq0$. As we will discuss
in the following, these two drawbacks are in fact related and are
simultaneously dealt with by extending noncommutative geometry into the
world of nonassociative geometry, which deals with deformations by
higher structures in geometry.

Another approach is based on string theory, which certainly provides a concrete
physical model of a quantum spacetime. Strings are extended
one-dimensional degrees of freedom and so, unlike the point particle
probes of quantum field theory, naturally come with intrinsic minimal
length $\Delta x^i\geqslant \ell_s$, where $\ell_s$ is the string
length. The interactions of strings thus violate locality, while the
theory directly contains gravity and is on-shell ultraviolet
finite. It is then natural to ask whether the spacetime approaches based on
noncommutative geometry and string theory are related or are complementary to
each other in some sense. The precise connection between \emph{open strings}
and noncommutative geometry was discovered near the end of the last
millenium, see e.g.~\cite{Douglas:1997fm,Ardalan:1998ks,Ardalan:1998ce,SheikhJabbari:1998ac,Chu:1998qz,SheikhJabbari:1999vm,Schomerus:1999ug,stncg3}, and is by now
well-established. Open strings on D-branes in background $B$-fields
(which provide a gauge flux on the D-brane worldvolume) probe a
noncommutative worldvolume geometry. The massless bosonic field
content of open string theory consists of gauge fields $A_i$ on the
worldvolume together
with scalar fields $\phi^a$ governing the transverse fluctuations of
the D-branes in spacetime. In the Seiberg-Witten scaling limit which
decouples open and closed string modes, the effective low-energy
dynamics is governed by a noncommutative gauge theory on the D-brane
worldvolume.

Although this connection is precise, and has led to a flurry of
investigation over the last 20 years, this does not explain the
connection of noncommutative geometry with gravity, nor how
noncommutative geometry can be used to formulate a consistent theory
of quantum gravity. In the present context, one needs to look at
\emph{closed strings}. The massless bosonic field content of closed
string theory consists of the spacetime metric $g_{ij}$, the
Kalb-Ramond field $B_{ij}$, and the dilaton field $\Phi$. Thus the closed string sector
contains the data of background geometry and gravity, and it is here
that one should seek analogs of a quantum geometry and suitable
decoupling limits to make contact with the problem of quantizing
gravity. Such a connection should certainly appear if noncommutative geometry
indeed improves the ultraviolet behaviour of quantum gravity.

The connections between noncommutative geometry and closed string
theory has been a topic of increasing interest over the last eight
years, where it has been realised that closed strings are related to
not only noncommutative but even \emph{nonassociative} target space
geometries. To understand how nonassociativity can arise in closed
string theory, it is helpful to take a step back and look at in more
detail why noncommutativity emerges in open string theory. An instructive 
pedagogical analogy from quantum mechanics is provided by the Landau
problem~\cite{Szabo:2004ic}: The planar quantum dynamics of electrons of mass $m$ and charge $e$ propagating
under the influence of a perpendicularly applied constant background magnetic
field of magnitude $B$. The Lagrangian is
\bea
L = \frac m2\, \dot{\vec x}{}^{\,2} - e\, \dot{\vec x}\cdot\vec A \qquad
\mbox{with} \quad A_i=-\frac B2\, \varepsilon_{ij}\, x^j \ .
\eea
The limit $B\gg m$ of strong magnetic field induces the projection
onto the lowest Landau level, described by a first order Lagrangian
$L\big|_{m=0} = \frac{e\,B}2\, \dot x^i\, \varepsilon_{ij}\, x^j$ with
degenerate phase space whose canonical quantization gives the commutation relations of a
noncommutative space:
\bea
[x^i,x^j] = \frac{\ii\hbar}{e\,B}\, \varepsilon^{ij} =: \ii\hbar\,
\theta^{ij} \ .
\eea

This simple model is analogous to that of bosonic open strings in a
$B$-field background, which at tree-level in string perturbation theory is described
generally by the
worldsheet action
\bea\label{eq:sigmamodel}
S = \frac1{4\pi\,\ell_s^2}\, \int_{\Sigma_2} \, 
\big(g_{ij}(x) \, \dd x^i\wedge\ast\, \dd x^j-2\pi\, \ell_s^2\, B_{ij}(x) \,\dd x^i\wedge \dd x^j \big) \ ,
\eea
where $x^i$ are maps from the string worldsheet, which is a
disk $\Sigma_2$, to the target spacetime
$M$. The low-energy limit
$\ell_s\to0$ describes the decoupling of massive string states, while the
Seiberg-Witten scaling limit $g_{ij}\sim\ell_s^4\to0$ ensures that 
gravity is non-dynamical and that the bulk modes of the string
decouple from the boundary degrees of freedom. In this limit the
action \eqref{eq:sigmamodel} reduces to a simple topological action given by the pullback of the Kalb-Ramond two-form to the string worldsheet:
\bea\label{eq:sigmamodelscaling}
S\big|_{g,\ell_s=0} = -\frac12\, \int_{\Sigma_2}\, B_{ij}(x)\, \dd x^i\wedge\dd x^j \ .
\eea
In the absence of NS--NS flux $H=\dd B=0$, and for target space
$M=\real^d$, using Stokes' theorem the action
\eqref{eq:sigmamodelscaling} gives a pure boundary interaction $S\big|_{g,\ell_s=0} = -\oint_{\partial\Sigma_2}\, \dd t \ \vec{\dot x}\cdot \vec A$, where $B=\dd A$ and $\dot x^i=\partial_tx^i$ denotes a tangential derivative of the
string field along the worldsheet boundary circle
$\partial\Sigma_2$. Again this action is of first order in worldsheet
time derivatives, so that the open string endpoints have a degenerate
phase space in the decoupling limit. For a constant $B$-field with the symmetric gauge choice $A_i = -\frac12\, B_{ij}\, x^j$, the action becomes
\bea
S\big|_{g,\ell_s=0} = \frac12\, \oint_{\partial\Sigma_2}\, \dd t \ x^i\,
B_{ij}\, \dot x{}^j \ ,
\eea
and if the $B$-field is moreover
non-degenerate then its canonical quantization produces the noncommutative coordinate algebra
\bea\label{eq:sigmamodelcomm}
[x^i,x^j] = \Big(\frac{\ii\hbar}B\Big)^{ij} =: \ii\hbar\, \theta^{ij} \ .
\eea
Thus the scaling limit of the open string sigma-model is formally
analogous to projection to the lowest Landau level for charged
particles in strong uniform magnetic fields.

The more precise dynamical mechanism behind this heuristic argument
can be infered from studying the two-point disk correlators deformed by the
non-zero two-form $B_{ij}$, which plays the role of a magnetic flux on
the worldvolume and can be turned on by a left-right asymmetric rotation
of the D-brane via T-duality. The $B$-field allows one to distinguish the 
insertions of string fields $x^i(t)$ and $x^j(t')$ on the boundary of
the disk in the correlation function, which depends only on the
ordering of the two boundary insertion points~\cite{stncg3}:
\bea\label{eq:openstringprop}
\big\langle x^i(t)\, x^j(t')\big\rangle = -\ell_s^2\, G^{ij}
\log(t-t')^2 + \frac{\ii\hbar}2\, \theta^{ij}\, {\rm sgn}(t-t') \ ,
\eea
where we used the open-closed string relations
\bea\label{eq:openclosedrel}
\frac1{g+2\pi\,\ell_s^2\, B} = \frac1G + \frac\theta{2\pi\, \ell_s^2} \ ,
\eea
with $G$ the open string metric and the bivector $\theta$ is the source of noncommutativity since
it is not symmetric under interchange of $x^i$ with $x^j$. The transformation \eqref{eq:openclosedrel} is familiar from the B\"uscher rules for T-duality, with the precise connection suggested by the worldsheet approach of~\cite{Duff}, and it explicitly determines the open string variables $(G,\theta)$ in terms of the closed string variables $(g,B)$ by
\bea
G= g-(2\pi\,\ell_s^2)^2\, B\, g^{-1}\, B \qquad \mbox{and} \qquad \theta=-(2\pi\,\ell_s^2)^2\, G^{-1}\, B\, g^{-1} \ .
\eea
Note that $G=g$ and $\theta=0$ exactly when $B=0$.
Using this
correlation function, the operator product expansion of open string
tachyon vertex operators on the boundary of the disk is computed to be
\bea\label{eq:boundaryOPE}
\e^{\ii k\cdot x(t)}\cdot\e^{\ii q\cdot x(t')} = (t-t')^{2\ell_s^2\,
  k_i\, G^{ij}\, q_j}\, \e^{-\frac{\ii\hbar}2\, k_i\, \theta^{ij}\,
q_j}\, \e^{\ii(k+q)\cdot x(t')} + \cdots \ ,
\eea
for $t>t'$.
The second factor in \eqref{eq:boundaryOPE} does not depend on the
worldsheet coordinates and is purely a target space effect, and in the
low-energy limit $\ell_s\to0$ whereby $\theta=B^{-1}$, this phase
factor is encoded in scattering amplitudes by the star product of fields $f,g$ given by
\bea
(f\star g)(x) = \int\, \dd k \ \int\, \dd q \ \tilde f(k)\, \tilde g(q) \, \e^{-\frac{\ii\hbar} 2\, k_i\,\theta^{ij}\, q_j} \, \e^{\ii (k+q)\cdot x}
\eea
in Fourier space, which is equivalent to the formal expansion in terms of a bidifferential operator:
\bea\label{eq:Moyal}
f\star g = \cdot\, \exp\Big(\frac{\ii\hbar}2\,
\theta^{ij}\, \partial_i\otimes\partial_j\Big)(f\otimes g) \ .
\eea
This is simply the Moyal-Weyl star product which is a noncommutative
deformation of the pointwise product $f\cdot g$ of functions on
spacetime. Its characteristic features are that it quantizes the
commutator $[x^i,x^j]=\ii\hbar\,\theta^{ij}$ (by defining
$[f,g]_\star:=f\star g- g\star f$ and setting $f=x^i$, $g=x^j$),
and it is associative: $f\star(g\star h) = (f\star g)\star h$. Since
$f\star g$ differs from $f\cdot g$ by a total derivative, the star
product deformation is consistent with the conformal $SL(2,\real)$
symmetry of the worldsheet theory which leaves the cyclic ordering of
boundary vertex operator insertions invariant, in the sense that the
star product is \emph{2-cyclic}:
\bea\label{eq:2cyclic}
\int\, \dd x \ f\star g = \int\, \dd x \ g\star f = \int\, \dd x \
f\cdot g \ .
\eea
Moreover, consistency with associativity of the operator product
expansion in conformal field theory only requires crossing symmetry of
the worldsheet correlation functions, which leads to the weaker
\emph{3-cyclic} condition:
\bea\label{eq:3cyclic}
\int\, \dd x \ f\star(g\star h) = \int\, \dd x \ (f\star g)\star h \ .
\eea
It is in this way that one arrives at a noncommutative gauge theory
for the massless bosonic open string modes $A_i$ and $\phi^a$; see e.g.~\cite{Douglas:2001ba,Szabo:2001kg} for reviews and further details.

Now let us try to understand how an analogous scenario could be realised in a
connection between noncommutative geometry and closed strings. Closed
strings see geometry in a different way than open strings do, which
from a target space perspective is due to T-duality. From the
worldsheet perspective, the
relevant tree-level amplitude involves correlation functions on the
sphere $S^2$, but the situation must be different and one has to
pass to higher correlators, as first pointed out by~\cite{stnag1}, since now the ordering on $S^2$ is
ambiguous because two points can be interchanged by an $SL(2,\real)$
transformation. However, the insertion of three string fields on
the sphere depends only on the relative orientation of the three
points, i.e. whether the insertion of a third point lies on the same
or opposite hemisphere as the other two points. A trivector flux
$\theta^{ijk}$ can be used to distinguish configurations, and in
analogy with the Moyal-Weyl star product \eqref{eq:Moyal} it deforms the algebra
of functions with the ``triproduct''
\bea\label{eq:triproduct}
f_1\tri f_2\tri f_3 = \cdot\,\exp\Big(\frac{\ii\hbar}6\,
\theta^{ijk}\, \partial_i\otimes\partial_j\otimes\partial_k\Big)(f_1\otimes
f_2\otimes f_3) \ ,
\eea
which leads to a \emph{nonassociative} tribracket defined by
\bea\label{eq:tribracketprod}
[f_1,f_2,f_3]_{\tri} := \sum_{\tau\in S_3}\, {\rm
  sgn}(\tau) \, f_{\tau(1)}\tri f_{\tau(2)}\tri f_{\tau(3)} \ .
\eea
This quantizes the basic coordinate brackets
\bea
[x^i,x^j,x^k]_{\tri} = \ii\hbar\, \theta^{ijk} \ .
\eea

The purpose of these lectures is to discuss, and answer as far as possible,
the following imminent questions at this stage:
\begin{itemize}
\item[{\bf (Q1)}] What is the trivector $\theta^{ijk}$?
\end{itemize}
We will see that this trivector is a `locally non-geometric flux',
called the $R$-flux. To properly discuss this, we shall have to review
some ingredients of non-geometric flux compactifications, generalized
geometry, and double field theory, some aspects of which are discussed in the lectures by C.~Hull at this School, and which we undertake in Section~\ref{sec:fluxes}. 
\begin{itemize}
\item[{\bf (Q2)}] What is the origin of the triproduct $\tri$?
\end{itemize}
We will see that the nonassociativity encoded in off-shell closed
string amplitudes is probed by suitable redefinitions of the
coordinate fields $x^i$ in linear flux backgrounds. We shall find in
Sections~\ref{sec:geometry} and~\ref{sec:quantization} that the triproduct is not the fundamental algebraic
entity, but arises as the result of a non-vanishing Jacobiator for a
\emph{nonassociative star product} on the closed string phase space.
\begin{itemize}
\item[{\bf (Q3)}] Is there a nonassociative version of the closed string effective
  action?
\end{itemize}
Recall that the closed string effective action for the massless
bosonic modes $g_{ij}$, $B_{ij}$ and $\Phi$ is given by
\bea\label{eq:Sgrav}
S_{\rm grav} = \frac1{16\pi\, G}\, \int_M\, \Big(\ast\, {\rm Ric} - \frac1{12}\,
\e^{-\Phi/3}\, H\wedge\ast\, H -
\frac16\, \dd\Phi\wedge\ast\, \dd\Phi \Big) \ .
\eea
Conformal invariance of the worldsheet theory at one-loop requires
vanishing beta-functions, which are equivalent to the target space equations of
motion resulting from \eqref{eq:Sgrav}. We shall discuss some aspects
of this far-reaching future goal in Section~\ref{sec:further}, but will not provide a complete and
decisive answer to the problem of the relevance of a nonassociative
theory of gravity in closed string theory, which is currently a topic
of ongoing investigation.

%%%%%%%%%%%%%%%%%%%%%%%%%%%%%%%%%%%%%%%%%%%%%%%%%%%%%%%
%%%%%%%%%%%%%%%%%%%%%%%%%%%%%%%%%%%%%%%%%%%%%%%%%%%%%%%

\section{A first glimpse at nonassociative geometry: Magnetic monopoles}
\label{sec:monopole}

As we saw in the case of open strings, a simple yet instructive
quantum mechanical analogue for the appearance of noncommutative
geometry is provided by the motion of electric charges in background
magnetic fields. A straightforward but far-reaching extension of this model
likewise provides an instructive physical scenario in which to understand the
appearence and implications of nonassociative geometry in the closed
string sector. We shall see later on that this model has a precise
analogue for closed strings propagating in locally non-geometric flux
backgrounds, and it enables us to introduce some of the geometric
ideas that will be used throughout this paper. The treatment of the quantum mechanical system of this
section is originally due to~\cite{Jackiw1985,Jackiw:2002wf}.

Consider the motion of a charged particle on $\real^3$ in a fixed
magnetic field $\vec B$, possibly with sources. The kinematical
momentum of the particle is $\vec p=m\,\dot{\vec x}$, which is the
physical gauge-invariant quantity and is not to be confused with the (gauge-variant)
canonical momentum. The Hamiltonian is taken to be the kinetic energy
\bea\label{eq:kineticHam}
H = \frac{\vec p{}^{\,2}}{2m} \ .
\eea
In the quantum theory, the Lorentz-Heisenberg equations of motion
\eq
\dot{\vec p} = \frac\ii\hbar\, [H,\vec p\,] = \frac e{2m}\, \big(\vec
p\times\vec B-\vec B\times\vec p\big) \qquad \mbox{and} \qquad
\dot{\vec x} = \frac\ii\hbar\, [H,\vec x\,] = \frac{\vec p}m
\eqend
require the \emph{deformed} canonical commutation relations of a noncommutative momentum space:
\bea\label{eq:monopolealgebra}
[x^i,x^j]=0 \ , \qquad [x^i,p_j]=\ii\hbar\, \delta^i{}_j \ , \qquad
[p_i,p_j] = \ii\hbar\, e\, F_{ij}(\vec x) \qquad \mbox{with} \quad
F_{ij} = \varepsilon_{ijk}\, B^k \ .
\eea
This formulation depends only on the magnetic field $\vec B$, and in
particular it allows for cases in which $\nabla\cdot\vec B\neq0$.

Let us understand the geometric structure underlying these commutation
relations. Writing phase space coordinates collectively as
$x^I=(x^i,p_i)$, we can express the relations in the form
\bea\label{eq:monopoleNC}
[x^I,x^J] = \ii\hbar\, \Theta^{IJ} \qquad \mbox{with} \quad (\Theta^{IJ}) =
\bigg(\begin{matrix} 0 & 1_3 \\ -1_3 & e\, F(\vec x) \end{matrix} \bigg) \
. 
\eea
The phase space bivector $\Theta=\frac12\, \Theta^{IJ}\, \partial_I\wedge\partial_J$ is
\emph{not} a Poisson bivector in general. The failure of the Jacobi
identity for the quasi-Poisson brackets defined by $\Theta$ is controlled
by the Schouten bracket
\bea
[\Theta,\Theta]_{\rm S}^{IJK} = \Theta^{[\underline{I}L}\, \partial_L\Theta^{\underline{JK}]}
\ ,
\eea
which is the natural extension of the usual Lie bracket of vector
fields to multivector fields; here only underlined indices are antisymmetrized. Then $[\Theta,\Theta]_{\rm S}=0$ if and only
if the Jacobiator
\bea
[x^I,x^J,x^K] := [x^I,[x^J,x^K]] + [x^J,[x^K,x^I]] +[x^K,[x^I,x^J]]
\eea
vanishes. An easy calculation shows
\bea
[p_1,p_2,p_3] = 3\,\hbar^2\, e\, \nabla\cdot\vec B =: 3\,\hbar^2\, e\,
\mu_0\, \rho_m \ .
\eea
Therefore the phase space algebra of the charged particle is
\emph{nonassociative} in the presence of magnetic sources
$\rho_m\neq0$.

For source-free magnetic fields $\vec B$, one has $\rho_m=0$ and
$\nabla\cdot\vec B=0$, so that there exists a globally defined
magnetic vector potential $\vec A$ on $\real^3$ such that $\vec B =
\nabla\times\vec A$. Then the commutation relations can be transformed
to canonical form with the canonical momentum $\vec\pi = \vec
p+e\,\vec A$. However, since $\vec B = \nabla\times\vec A$ if and only if
$\nabla\cdot\vec B=0$, we cannot work with canonical momenta and
covariant derivatives in the presence of magnetic sources, i.e. for
$\rho_m\neq0$ we encounter nonassociativity and there is no linear
operator $\vec p=\ii\hbar\, \nabla-e\,\vec A$. Let us now explore how
to understand magnetic sources and the ensuing violation of the Jacobi
identity.

Since $[x^i,p_j]=\ii\hbar\,\delta^i{}_j$, translations in the quantum
theory are generated by the magnetic translation operators
\bea\label{eq:magnetictranslations}
T(\vec a) = \e^{\frac\ii\hbar\, \vec a\cdot\vec p}
\eea
with $T^{-1}(\vec a)\,\vec x\,T(\vec a) = \vec x+\vec a$. These
operators do not form a representation of the translation group on
$\real^3$, as a simple calculation shows
\bea
T(\vec a_1)\, T(\vec a_2) = \e^{\frac{\ii e}\hbar\, \Phi_2(\vec
x;\vec a_1,\vec a_2)} \ T(\vec a_1+\vec a_2) \ ,
\eea
where
\bea
\Phi_2(\vec x;\vec a_1,\vec a_2) = \int_{\langle\vec a_1,\vec
  a_2\rangle_{\vec x}} \, \vec B\cdot\dd\vec S
\eea
is the magnetic flux through the oriented triangle $\langle\vec
a_1,\vec a_2\rangle_{\vec x}$ based at $\vec x\in\real^3$ with sides
$\vec a_1$, $\vec a_2$ and $\vec a_1+\vec a_2$. The Jacobi identity is
the infinitesimal statement of associativity, and its failure results
in the relations
\bea
\big(T(\vec a_1)\, T(\vec a_2)\big)\, T(\vec a_3) = \e^{\frac{\ii
    e}\hbar\, \Phi_3(\vec x;\vec a_1,\vec a_2,\vec a_3)} \ T(\vec
a_1)\, \big(T(\vec a_2)\, T(\vec a_3)\big) \ ,
\eea
where
\bea
\Phi_3(\vec x;\vec a_1,\vec a_2,\vec a_3) = \int_{\partial\langle\vec
  a_1,\vec a_2,\vec a_3\rangle_{\vec x}}\, \vec B\cdot\dd\vec S =
\int_{\langle\vec a_1,\vec a_2,\vec a_3\rangle_{\vec x}}\,
\nabla\cdot\vec B \ \dd V
\eea
is the magnetic charge $q_m$ enclosed by the oriented tetrahedron $\langle\vec
a_1,\vec a_2,\vec a_3\rangle_{\vec x}$ based at $\vec x\in \real^3$
with sides $\vec a_1$, $\vec a_2$, $\vec a_3$, $\vec a_1+\vec a_2$,
$\vec a_2+\vec a_3$ and $\vec a_1+\vec a_2+\vec a_3$, and we have used
the divergence theorem.

It follows that associativity of translations is ensured 
when~\cite{Jackiw1985}
\bea
\frac{\mu_0\,e\,q_m}\hbar \in 2\pi\,\zed
\eea
which is the celebrated Dirac charge quantization condition. Then the
usual quantum mechanical formalism can be applied with linear
operators on a separable Hilbert space. But this restricts the form of
the magnetic field $\vec B$, which must be sourced by a point-like
magnetic monopole (or a collection thereof), so that the phase
$\Phi_3$ does not lose its integrality when the translation vectors
$\vec a_i$ are continuously varied. In this case the magnetic source
must lie either inside or outside the tetrahedron $\langle\vec
a_1,\vec a_2,\vec a_3\rangle_{\vec x}$, and is given by the Dirac
monopole field
\bea\label{eq:Diracfield}
\vec B = \mu_0\, q_m\, \frac{\vec x}{|\vec x|^3} \qquad \mbox{with}
\quad \nabla\cdot\vec B = 4\pi\, \mu_0\, q_m\, \delta(\vec x) \ .
\eea

What becomes of the Jacobi identity in this case? We note that it is
violated precisely at the loci of the magnetic charges, which for
Dirac monopoles occur at isolated points and so can be excised from
$\real^3$, where the magnetic field $\vec B$ is singular. Such an
excision is also natural from the point of view of angular momentum
conservation, which imples that the electric charges never reach the
monopoles and their wavefunction vanishes at the monopole
locations~\cite{BL}. This leads to a geometric description of Dirac monopoles in
terms of connections on a non-trivial $U(1)$-bundle
$P\to\real^3\setminus\{\vec 0\}\simeq S^2$ of first Chern class
$c_1(P)=\mu_0\,e\,q_m/2\pi\,\hbar$, and the wavefunctions of the particle live in the Hilbert space of square-integrable sections of $P$. In this case the map $\vec
a\mapsto T(\vec a)$ defines a projective representation of the
translation group of $\real^3\setminus\{\vec 0\}$ on this Hilbert space, and the projective
phase $\e^{\frac{\ii e}\hbar\, \Phi_2}$ is the group two-cocycle of the representation.

For our later considerations we are interested in situations corresponding to a
{constant} homogeneous magnetic charge density background $\rho_m$,
whose algebraic structure was first studied in~\cite{Gunaydin:1985ur}. The analogue of the
rotationally symmetric field \eqref{eq:Diracfield} in this case is
given by
\bea\label{eq:linearmagfield}
\vec B = \frac{\mu_0\,\rho_m}3\, \vec x \ .
\eea
The magnetic charge is now uniformly distributed over all space, so
that the phase space coordinate algebra becomes everywhere
nonassociative. In this case removing the magnetic sources from
$\real^3$ would leave an empty space. In this sense the momentum space
of an electric charge in a uniform magnetic charge distribution is
`locally non-geometric'.  The constant magnetic charge density is not
described by a connection on a $U(1)$-bundle over
$\real^3\setminus\{\vec 0\}$, but rather in terms of a connection on a
(trivial) $U(1)$-gerbe on $\real^3$, i.e. by a ``$B$-field''
\bea
F_{ij} = \varepsilon_{ijk}\, B^k = \frac{\mu_0\,\rho_m}3\,
\varepsilon_{ijk}\, x^k
\eea
with curvature $H=\dd F=\mu_0\,\rho_m\, \dd x^1\wedge\dd x^2\wedge \dd
x^3$; since $H\neq0$ everywhere there is not even a local magnetic vector
potential $\vec A$ in this instance. Moreover, now the phase factors $\Phi_2$
yield two-cochains, rather than two-cocycles, whose coboundary is the phase
$\e^{\frac{\ii e}\hbar\, \Phi_3}$ which hence defines a three-cocycle of the
translation group of $\real^3$. Thus the quantum theory with this
three-cocycle nessecitates \emph{nonassociative quantum
mechanics}. Geometrically, the bivector $\Theta$ in this case defines a
`twisted Poisson structure' on phase space. As
we will see later on, such a quantum system makes perfect physical sense and possesses fascinating properties.

This simple quantum mechanical example illustrates the string
theoretical considerations which will follow, in the topic of
non-geometric flux compactifications that we turn to next. Our string
theory considerations will lead naturally to an approach to
nonassociative geometry via deformation quantization, analogously to
the open string case, which will be the main tool of this
paper. However, as phase space quantum mechanics comes with its own
issues, as we discuss later, let us point out for completeness some
alternative approaches to the nonassociative quantum mechanics alluded
to above, of which there are currently three that each have their own
deficiencies as well. 

Firstly, and most straightforwardly, one may generalize the technique of symplectic realization from Poisson geometry to the twisted Poisson structure $\Theta$ and embed the nonassociative phase space into a symplectic manifold of twice the original dimension~\cite{Kupriyanov:2018xji}. In this associative framework standard techniques of geometric or canonical quantization are available, and in particular a global magnetic vector potential exists on the doubled configuration space; its drawback is that, while the doubling is tantalizing reminescent of the framework of double field theory discussed below, it is not possible to eliminate the spurious auxiliary degrees of freedom that enable the reformulation in terms of associative geometry. Secondly, one can exploit the geometric structure of the gerbe associated to the distribution of magnetic charge to face the nonassociativity head on and define quantum states that live in the 2-Hilbert space of sections of this gerbe, in analogy to the ordinary Hilbert space of sections of a line bundle in the source-free case~\cite{Szabo:2017yxd,Bunk18}. This gives a geometric description of the magnetic translation operators \eqref{eq:magnetictranslations} acting on this 2-Hilbert space with the three-cocycle above interpreted as a higher projective phase of a 2-representation; the drawback of this approach is that it is rather technically complicated and it is difficult to represent observables such as the Hamiltonian operator \eqref{eq:kineticHam} on the 2-Hilbert space, while conceptually it is not clear what is the meaning of such higher quantum states. Thirdly, one can apply transgression techniques to map the gerbe to a line bundle over the loop space of the configuration manifold~\cite{Saemann2011,Saemann13,Bunk16}. This approach is naturally suggested by the closed string origin of nonassociativity, and with it one can apply standard techniques of geometric quantization on loop space which successfully captures some predictions of string theory; the drawback of this approach is that it requires difficult infinite-dimensional analysis which makes computations of physical quantities, such as expectation values, seemingly intractable. Hence in all approaches one trades nonassociativity for some other sort of technical or conceptual complication.

%%%%%%%%%%%%%%%%%%%%%%%%%%%%%%%%%%%%%%%%%%%%%%%%%%%%%%%
%%%%%%%%%%%%%%%%%%%%%%%%%%%%%%%%%%%%%%%%%%%%%%%%%%%%%%%

\section{Non-geometric fluxes and nonassociative geometry}
\label{sec:fluxes}

The purpose of this section is to explain what the notion of ``non-geometry'' means in string theory, and in particular to answer Question (Q1) from Section~\ref{sec:intro}. Recall that in closed string theory, in addition to the spacetime metric $g_{ij}$, there is a massless Kalb-Ramond two-form $B$ in the NS--NS sector with curvature three-form $H=\dd B$. The motion of the strings is described by the two-dimensional non-linear sigma-model action \eqref{eq:sigmamodel}. The classical vacua are described by two-dimensional conformal field theories, and in this setting the target space geometry is emergent. On the other hand, there exist conformal field theories which cannot be identified with such simple large radius geometries, such as left-right asymmetric orbifolds, and in such instances we will advocate the point of view that the target space interpretation is related to noncommutative and nonassociative geometry. This is the case when a gauge flux is turned on in the worldvolume field theory of a D-brane, which corresponds to an asymmetric rotation in the boundary conformal field theory.

Another left-right asymmetric worldsheet operation is that of
T-duality, which reflects right-moving strings while leaving the
left-moving sector unchanged. From the perspective of a
$d$-dimensional target space, a T-duality $\Tcal_i$ along the $i$-th
direction exchanges momentum modes $p_i$ with winding modes $w^i$, and
correspondingly for the canonically conjugate variables: position
coordinates $x^i$ are exchanged with their dual ``winding
coordinates'' $\tilde x_i$. The collection of symmetry transformations
form the split-signature orthogonal group $O(d,d)$ which is the continuous extension of the physical T-duality group
$O(d,d;\zed)$ of toroidally compactified closed string theory.

String theory with fluxes is of interest both for its geometric allure
and because of its relevance to observable phenomenology and
cosmology (see e.g.~\cite{Grana,Douglasflux,BKLSrev} for reviews): Flux compactifications can lead to generalized geometric
structures, obtained for example by patching together with string
symmetries, while at the same time they stabilize moduli on the
string landscape. They are also of importance in the AdS/CFT
correspondence. Starting from flat space with non-vanishing NS--NS $H$-flux, T-duality gives rise to a chain of geometric and non-geometric fluxes~\cite{Hull:2004in,Shelton:2005cf}
\bea\label{eq:Tdualitychain}
H_{ijk}\xrightarrow{ \ \Tcal_i \ } f^i{}_{jk}\xrightarrow{ \ \Tcal_j \ }Q^{ij}{}_k\xrightarrow{ \ \Tcal_k \ } R^{ijk} \ .
\eea
Let us briefly describe the geometrical meaning of each member of this
T-duality chain. The first term is of course the geometric NS--NS flux $H=\dd B$, which represents the characteristic class of a $U(1)$-gerbe by generalized Dirac quantization of fluxes in string theory. The second term is a metric flux, which appears as torsion in the geometry: In a suitable basis of vielbeins $e^i$, with inverses $e_i$, it appears in the Maurer-Cartan equations $\dd e^i = -\frac12\, f^i{}_{jk}\, e^j\wedge e^k$, or equivalently as the structure constants of the non-trivial Lie bracket of vector fields $[e_i,e_j]=f^k{}_{ij}\, e_k$. The third member is the first example of a non-geometric frame, and $Q^{ij}{}_k$ is called a `globally non-geometric' $Q$-flux. These are also called T-folds, which have a local description in terms of conventional Riemannian geometry, but globally the transition functions between local charts also involve T-duality transformations. The final member of the chain is the most mysterious frame and the one which shall occupy most of our attention: Here $R^{ijk}$ is called a `locally non-geometric' $R$-flux. This background cannot even be described locally by conventional geometry and has no clear target space interpretation. As we discuss below it is this frame that this gives rise to a \emph{nonassociative geometry}.

Describing these fluxes and understanding this non-geometric regime of
string theory requires generalized geometry and its extension to double field theory,
which is an $O(d,d)$-symmetric theory treating the metric and
$B$-field on equal footing, and which is covered in the lectures by C.~Hull at this School. 

\subsection{Generalized geometry}

The geometric and non-geometric
fluxes appearing in the T-duality chain \eqref{eq:Tdualitychain} can
be regarded as structure constants of a generalized bracket
\eq\label{eq:Roytenberg}
\begin{aligned}
{}[e_i,e_j]_{\rm R} &= f^k{}_{ij}\, e_k + H_{ijk}\, e^k \ , \\[4pt]
[e^i,e_j]_{\rm R} &= f^i{}_{jk}\, e^k - Q^{ik}{}_j\, e_k \ , \\[4pt]
[e^i,e^j]_{\rm R} &= Q^{ij}{}_k\, e^k + R^{ijk}\, e_k  \ ,
\end{aligned}
\eqend
for a local vielbein basis of sections $(e_i,e^i)$ of the vector
bundle $E=TM\oplus T^*M$, which in generalized geometry is called the \emph{generalized
  tangent bundle}~\cite{Hitchin:2004ut,Gualtieri:2003dx}. In this form the bracket is usually
called the Roytenberg bracket~\cite{RoytenbergContemp,Halmagyi,BlumenhagenR}, and its reductions for various choices
of vanishing fluxes gives the usual Courant and Dorfman brackets of
generalized geometry. It governs the worldsheet current
algebras in flux compactifications of string theory, see
e.g.~\cite{Alekseev:2004np,Halmagyi:2008dr,Bouwknegt:2010zz}.

The sections of the generalized tangent bundle $E=TM\oplus T^*M$ are denoted as $X+\xi$ with $X = X^i\,
e_i$ a
vector field and $\xi = \xi_i\, e^i$ a one-form on $M$. The bundle $E$ carries a canonical
$O(d,d)$-structure through the natural pairing $\langle e_i,e^j\rangle
= \delta_i{}^j$ of the tangent bundle $TM$ with the cotangent bundle
$T^*M$ of the target space $M$. The structure group $O(d,d)$ has two
natural abelian subgroups acting on sections of the generalized
tangent bundle in the following way:
\begin{itemize}
\item \underline{$B$-transforms:} \ \ 
    $\displaystyle{\bigg( \begin{matrix} 1_d & 0 \\ B &
      1_d \end{matrix} \bigg)} \ : \ X+\xi\longmapsto X + \big(\xi+ \iota_XB\big)
  \ , $
\item \underline{$\theta$-transforms:} \ \ 
  ${\displaystyle{\bigg( \begin{matrix} 1_d & \theta \\ 0
    & 1_d \end{matrix} \bigg)}} \ : \ X+\xi\longmapsto
\big(X+\theta^\sharp\xi\big) + \xi \ , $
\end{itemize}
where $B$ is a two-form and $\theta$ is a bivector which induce the natural contraction
maps $\iota:TM\to T^*M$, $(\iota_XB)_i = B_{ij}\, X^j$ and
$\theta^\sharp: T^*M\to TM$, $(\theta^\sharp\xi)^i = \theta^{ij}\,
\xi_j$. Any $O(d,d)$-transformation $\co\in O(d,d)$ can be written in the form
\bea
\co = \bigg( \begin{matrix} 1_d & 0 \\  B &
      1_d \end{matrix} \bigg) \, \bigg( \begin{matrix} N & 0 \\ 0 & N^{-\top} \end{matrix} \bigg) \, \bigg( \begin{matrix} 1_d & \theta \\ 0
    & 1_d \end{matrix} \bigg) \quad \mbox{with} \ \ \bigg( \begin{matrix} N & 0 \\ 0 & N^{-\top} \end{matrix} \bigg) \ : \ X+\xi\longmapsto \iota_XN + \big(N^{-\top}\big)^\sharp\xi
\eea
where $N\in GL(d)$ determines a general linear transformation of sections.

The generalization of the Lie bracket of vector fields $X,Y$ on $TM$ to sections $X+\xi,Y+\eta$ of $TM\oplus T^*M$ is provided by the Dorfman bracket
\bea
[X+\xi,Y+\eta]_{\rm D} = [X,Y] + \pounds_X\eta - \iota_Y\,\dd\xi \ ,
\eea
where $\pounds_X$ denotes the Lie derivative along $X$. Fluxes can be incorporated into this bracket structure by adding appropriate twisting terms, e.g. $\iota_X\iota_YH$. Alternatively, they may be added by applying suitable $O(d,d)$-transformations of sections of the generalized tangent bundle, starting from the standard geometric frame with metric flux, basis $(e_i,e^i)$ and the Dorfman brackets
\bea\label{eq:Dorfman}
[e_i,e_j]_{\rm D} = f^k{}_{ij}\, e_k \ , \qquad [e^i,e_j]_{\rm D} =
f^i{}_{jk}\, e^k \qquad \mbox{and} \qquad [e^i,e^j]_{\rm D} = 0 \ .
\eea
Then under a $B$-transform of the basis $(e_i,e^i)$ these brackets map into
\bea\label{eq:HtwistedDorfman}
[e_i,e_j]_{\rm R} = f^k{}_{ij}\, e_k + H_{ijk}\, e^k \ , \qquad
[e^i,e_j]_{\rm R} = f^i{}_{jk}\, e^k \qquad \mbox{and} \qquad
[e^i,e^j]_{\rm R} = 0 \ ,
\eea
with the geometric NS--NS flux $H=\dd B$, while under a $\theta$-transform they map to
\bea\label{eq:RtwistedDorfman}
[e_i,e_j]_{\rm R} = f^k{}_{ij}\, e_k \ , \quad [e^i,e_j]_{\rm R} = f^i{}_{jk}\, e^k - Q^{ik}{}_j\, e_k \quad \mbox{and} \quad 
[e^i,e^j]_{\rm R} = Q^{ij}{}_k\, e^k + R^{ijk}\, e_k \ ,
\eea
with the globally and locally non-geometric fluxes
\bea\label{eq:QRtheta}
Q^{ij}{}_k = \partial_k\theta^{ij} + 2\,f^{[\underline{i}}{}_{kl}\, \theta^{l\underline{j}]} \qquad \mbox{and} \qquad R^{ijk} = 3\,\big( [\theta,\theta]_{\rm S}^{ijk} + f^{[\underline{i}}{}_{lm}\, \theta^{\underline{j}m}\, \theta^{\underline{k}m]}\big) \ .
\eea
In this way the local $O(d,d)$-transformations of the Dorfman bracket \eqref{eq:Dorfman} reproduce the Roytenberg bracket \eqref{eq:Roytenberg}. What is particularly noteworthy and relevant for us here is that the non-geometric fluxes are determined by a bivector $\theta$.

\subsection{Double field theory}

Let us now briefly explain how double field theory~\cite{Siegel1,Siegel2,HullZwiebach} provides a
\emph{microscopic} description of $Q$-flux and $R$-flux, through the
chain \eqref{eq:Tdualitychain} of T-duality transformations, by a
formal definition and unified description of non-geometric fluxes; see e.g.~\cite{dftrev1,dftrev2,dftrev3} for
reviews. The idea of double field theory is to double the target space coordinates $x^i$ to $x^I=(x^i,\tilde x_i)$, where $\tilde x_i$ are the T-dual ``winding coordinates''. This gives a formalism with manifest $O(d,d)$-symmetry that allows one to perform such T-dualities to non-geometric frames. 

Double field theory is a field theory for the massless modes of closed bosonic string theory that treats diffeomorphism symmetry and $B$-field gauge transformations on equal footing by assembling them into the \emph{generalized metric}
\bea
\mcH = \bigg( \begin{matrix} g^{-1} & -2\pi\,\ell_s^2\, g^{-1}\, B \\ 2\pi\,\ell_s^2\,B\, g^{-1} & g- (2\pi\,\ell_s^2)^2\,B\, g^{-1}\, B \end{matrix} \bigg) \ .
\eea
This metric can be written in terms of a Schur decomposition
\bea
\mcH = \bigg( \begin{matrix} 1_d & 0 \\ 2\pi\,\ell_s^2\,B & 1_d \end{matrix}\bigg)\, \bigg( \begin{matrix} g^{-1} & 0 \\ 0 & g \end{matrix} \bigg)\, \bigg( \begin{matrix} 1_d & -2\pi\,\ell_s^2\,B \\ 0 & 1_d \end{matrix} \bigg) \ ,
\eea
which identifies it as a $B$-transform of the doubled space metric when $B=0$. There is a global $O(d,d)$-symmetry that includes T-duality and acts as
\bea
x^I\longmapsto \co^I{}_J\, x^J \ , \qquad \mcH\longmapsto \co^\top\, \mcH\, \co \qquad \mbox{with} \quad \co=(\co^I{}_J)\in O(d,d) \ .
\eea
One then constructs an $O(d,d)$-invariant action for $\mcH$ and halves
the degrees of freedom by imposing the strong constraint
$\partial_i\otimes\tilde\partial{}^i+
\tilde\partial{}^i\otimes\partial_i=0$ on all products of fields of double field theory, which is a
strong version of the level-matching condition $L_0-\overline{L}_0=0$
in the worldsheet conformal field theory, i.e. $p_i\, w^i=0$. Solving the strong constraint amounts to choosing a polarization on the doubled space~\cite{HullReidEdwards}; for example, in the geometric supergravity frame one takes $\tilde \partial{}^i:=\frac{\partial}{\partial\tilde x_i}=0$, which reduces the fields of double field theory to fields of generalized geometry. Different polarisations define different T-duality frames, and any two frames are related by an $O(d,d)$-transformation.

Let us consider the example of flat space, $g^\circ=1_d$, with constant
$H$-flux, and choose the Kalb-Ramond field in the symmetric gauge
$B^\circ_{ij}=\frac13\, H_{ijk}\, x^k$; this is locally defined (for
$x\in\real^d$), but not globally if the spacetime is e.g. a torus, where it is only defined up to large gauge transformations. A T-duality in the $i$-th direction interchanges $x^i$ with $\tilde x_i$ and is implemented by the factorized $O(d,d)$-transformation matrix
\bea
\Tcal_i = \bigg( \begin{matrix} 1_d-E_i & E_i \\ E_i & 1_d-E_i \end{matrix} \bigg) \ ,
\eea
where $E_i$ are the $d\times d$ matrix units $(E_i)_{kl} = \delta_{ki}\, \delta_{li}$. Consider now the effect of applying two successive T-duality transformations $\Tcal_{(ij)}:= \Tcal_i\, \Tcal_j$ to the corresponding generalized metric:
\bea
\begin{aligned}
\mcH_{(ij)} &= \Tcal_{(ij)}^\top \, \bigg( \begin{matrix} 1_d & -2\pi\,\ell_s^2\, B^\circ \\ 2\pi\,\ell_s^2\,B^\circ & 1_d- (2\pi\,\ell_s^2)^2\,(B^\circ)^2 \end{matrix} \bigg) \, \Tcal_{(ij)} \\[4pt] &=:
\bigg( \begin{matrix} g^{-1} & -2\pi\,\ell_s^2\, g^{-1}\, B \\
  2\pi\,\ell_s^2 \, B\, g^{-1} & g- (2\pi\,\ell_s^2)^2 \, B\, g^{-1}\, B \end{matrix} \bigg) \ .
\end{aligned}
\eea
One easily computes that the new metric and Kalb-Ramond field $(g,B)$ are not globally defined in the directions orthogonal to the $(x^i,x^j)$-plane, which is the earmark of the `global non-geometry' of the T-fold in the geometric parameterization. However, a suitable field redefinition appropriate to the transformation from a geometric frame to a non-geometric frame yields a new parameterization of the generalized metric in double field theory as~\cite{Andriot:2012wx}
\bea
\mcH_{(ij)} = \bigg( \begin{matrix} G^{-1} - \frac1{(2\pi\,\ell_s^2)^2}\, \theta\, G\, \theta & \frac1{2\pi\,\ell_s^2}\, \theta\, G \\ -\frac1{2\pi\,\ell_s^2}\, G\, \theta & G
\end{matrix} \bigg) \ ,
\eea
where the new metric $G$ and bivector $\theta=\frac12\,
\theta^{ij}\, \partial_i\wedge\partial_j$ are given precisely by the
open-closed string relation \eqref{eq:openclosedrel}, as anticipated
from the B\"uscher rules. This metric can similarly be obtained from a
$\theta$-transform given by
\bea
\mcH_{(ij)} = \bigg( \begin{matrix} 1_d & \frac1{2\pi\,\ell_s^2}\, \theta \\0 &
  1_d \end{matrix} \bigg)\, \bigg( \begin{matrix} G^{-1} & 0 \\ 0 &
  G \end{matrix} \bigg)\, \bigg( \begin{matrix} 1_d & 0 \\
  -\frac1{2\pi\,\ell_s^2}\, \theta &
  1_d \end{matrix} \bigg) \ .
\eea
In the present case one computes
\bea\label{eq:Qfluxframe}
G=1_d \qquad \mbox{and} \qquad \theta^{ij}=Q^{ij}{}_k\, x^k \ , 
\eea
which defines flat space with constant non-geometric $Q$-flux
\bea
Q^{ij}{}_k = \partial_k\, \theta^{ij} \ .
\eea

Finally, let us apply a T-duality transformation $\Tcal_k\in O(d,d)$ along a remaining non-isometric direction, which exchanges the physical coordinate $x^k$ with the winding coordinate $\tilde x_k$ and maps the geometric data \eqref{eq:Qfluxframe} to
\bea\label{eq:Rfluxframe}
\tilde G = 1_d \qquad \mbox{and} \qquad \tilde\theta^{ij} = R^{ijk}\, \tilde x_k \ .
\eea
This defines flat space with constant non-geometric $R$-flux
\bea\label{eq:Rfluxtheta}
R^{ijk} = \tilde\partial{}^{[i}\tilde\theta^{jk]} \ .
\eea
As the $R$-flux explicitly involves derivatives of the winding coordinates, it cannot be described in ordinary geometry and in this sense the $R$-flux frame is `locally non-geometric'. In this simple example, one can alternatively regard $\tilde\theta$ as a two-form on the dual winding space of curvature $R=\dd\tilde\theta$.

\subsection{Overview of conformal field theory results}

Let us now summarise the explicit worldsheet calculations which have suggested the appearance of noncommutative and nonassociative deformations of geometry in the non-geometric frames of closed string theory:
\begin{itemize}
\item The original suggestion of nonassociativity by~\cite{stnag1} computes the cyclic equal time double commutator 
\bea
[x^i,x^j,x^k]:=\lim_{\sigma_i\to\sigma} \ [[x^i(\sigma_1,\tau),x^j(\sigma_2,\tau)],x^k(\sigma_3,\tau)] + \mbox{cyclic}
\eea
in the $SU(2)$ Wess-Zumino-Witten model with $H$-flux, and finds a non-vanishing target space quantity.
\item Explicit calculations of phase space commutators by canonical
  quantization of closed strings in flat space and in a linear
  $B$-field background were carried out
  in~\cite{stnag2,Condeescu:2012sp,Andriot:2012vb}, by studying
  monodromy properties and the corresponding twisted closed string boundary
  conditions, which lead to a shifted closed string mode
  expansion analogous to the open string case and expansions in
  asymmetric orbifold string theories. These calculations reveal
  generally a doubled phase space
  nonassociative geometry in the different T-duality frames.
\item Correlators of products of tachyon vertex operators in
  sigma-model perturbation theory about a flat geometry with small constant $H$-flux were computed in~\cite{Blumenhagen2011}, and after conformal field theory T-duality shown to reproduce the triproducts discussed in Section~\ref{sec:intro}.
\end{itemize}
Further evidence is provided in e.g.~\cite{Chatzistavrakidis2013,Davidovic2013,Blair2014,Bakas:2015gia,Nikolic2018}.

The resulting quantum geometry structures can be described as follows. Let us first consider the $Q$-flux frame. Naively, in analogy with the open string case, the bivector of \eqref{eq:Qfluxframe} in the non-geometric parameterization would suggest noncommutativity, but in the closed string case one needs more: Only closed strings which wind in the $Q$-flux frame can probe a quantum deformation of the geometry, and the noncommutativity in this case is determined by a Wilson line of the $Q$-flux as~\cite{Andriot:2012an}
\bea
[x^i,x^j] = \frac{\ii\ell_s^4}{3\hbar}\, \oint\, \dd\theta^{ij} = \frac{\ii\ell_s^4}{3\hbar}\, \oint\, Q^{ij}{}_k\, \dd x^k = \frac{\ii\ell_s^4}{3\hbar}\, Q^{ij}{}_k\, w^k \ ,
\eea
where $w^k$ are the closed string winding numbers. Since the winding numbers are central elements in this algebra, i.e. $[w^i,w^j]=[x^i,w^j]=0$, these relations define a noncommutative but associative geometry.

To probe nonassociativity one needs to introduce local $R$-flux. The explicit computations above yield the tribracket
\bea\label{eq:tribracket}
[x^i,x^j,x^k]=\ell_s^4\, R^{ijk} \ .
\eea
Let us quickly sketch how this bracket is derived from conformal
perturbation theory, refering to the
original work~\cite{Blumenhagen2011} and also to the nice
review~\cite{Blumenhagen:2014sba} for details of the
calculation. Using complex coordinates on the Riemann sphere
$S^2\simeq\complex\cup\{\infty\}$, the worldsheet equations of motion
to linear order in the $H$-flux for a flat target space read
$\partial\overline{\partial} x^i= \frac12\, H^i{}_{jk}\,\partial
x^j\,\overline{\partial} x^k$, and thus the coordinate fields have to
be modified in order to be consistent with the conformal field theory
description, wherein the perturbation of the free worldsheet
sigma-model by the $H$-flux should yield a marginal deformation. We
therefore replace the usual conserved currents by
\eq
\CJ^i = \ii\partial x^i-\frac\ii2\, H^{i}{}_{jk}\, \partial x^j\, x_{\rm
  r}^k \qquad \mbox{and} \qquad
\overline{\CJ}{}^{\,i}=\ii\overline{\partial} x^i -\frac\ii2\,
H^{i}{}_{jk}\, x_{\rm l}^j\, \overline{\partial} x^k
\eqend
where $x^i=x_{\rm l}^i+x_{\rm r}^i$ is the decomposition of the string
fields into left and right moving modes; the dual winding fields are
then given by $\tilde x{}^i=x_{\rm l}^i-x_{\rm r}^i$. The correlation functions of
three insertions of the currents $\CJ^i$ are readily computed, and
after writing $\CJ^i =: \ii\partial\CX^i$ and performing three
worldsheet integrations, one arrives at the closed string
generalization of the second term in the open string propagator
\eqref{eq:openstringprop} for the modified string fields $\CX^i$ with
the sign function, that arises from combinations of the complex logarithm function, replaced by certain
combinations of the complex Rogers dilogarithm function, and the
bivector $\theta^{ij}$ substituted by the trivector $\theta^{ijk} =
\frac{\ell_s^4}{\hbar}\, H^{ijk}$.

A triple T-duality transformation $\Tcal_{(ijk)}:=\Tcal_i\,\Tcal_j\,
\Tcal_k$ is affected in the worldsheet conformal field theory as an
asymmetric reflection of the right-moving string coordinates: $\CX_{\rm
  l}^i\mapsto\CX_{\rm l}^i$, $\CX_{\rm r}^i\mapsto -\CX_{\rm r}^i$,
which maps winding modes $w^i$ in the $H$-flux frame to momentum modes $p_i$
in the $R$-flux frame. The three-point correlators of the
corresponding tachyon vertex operators $\e^{\ii k\cdot\CX}$ thereby
produce a phase which is trivial in the $H$-flux frame, but which is
non-trivial in the $R$-flux frame and encoded in scattering amplitudes
\bea
(f\tri g \tri h)(x) = \int\, \dd k \ \int\, \dd q \ \int\, \dd r \ \tilde f(k)\, \tilde g(q) \, \tilde h(r) \, \e^{-\frac{\ii\ell_s^4}{6}\, R^{ijk} \, k_i\, q_j\, r_k} \, \e^{\ii(k+q+r)\cdot x}
\eea
by the triproducts \eqref{eq:triproduct}
with $\theta^{ijk}=\frac{\ell_s^4}{\hbar}\, R^{ijk}$. These
phases are consistent with the crossing symmetry of conformal field
theory correlation functions, or equivalently associativity of the
operator product expansion, because by momentum conservation one has
$p_i\, q_j\, r_k\, \theta^{ijk}=0$ whenever $p+q+r=0$. This is
equivalent to the cyclicity property of the triproduct
\bea
\int\, \dd x\ f\tri g\tri h = \int\, \dd x\ f\cdot g\cdot h \ ,
\eea
which for constant $R$-flux follows from integration by parts since the two integrands differ by total derivatives.
Hence the triproducts are consistent with the axioms of conformal
field theory and nonassociativity is not probed by the on-shell
theory. Note that it is the modified string coordinates $\CX^i$ and
not the original ones $x^i$ which probe the nonassociative geometry, a
feature which is also confirmed in other worldsheet approaches~\cite{Bakas:2015gia}.

Alternatively, following what we did for the $Q$-flux frame, using the bivector from \eqref{eq:Rfluxframe} we find noncommutativity probed by closed strings which propagate in the $R$-flux frame given by
\bea
[x^i,x^j] = \frac{\ii\ell_s^4}{3\hbar}\, \oint\, \dd\tilde\theta^{ij} = \frac{\ii\ell_s^4}{3\hbar}\, \oint\, R^{ijk}\, \dd \tilde x_k = \frac{\ii\ell_s^4}{3\hbar}\, R^{ijk}\, p_k
\eea
where $p_k$ are the closed string momentum modes. However, now the variables $x^i$ and $p_i$ are canonically conjugate, and so the tribracket \eqref{eq:tribracket} can be regarded as the Jacobiator of the precursor nonassociative phase space algebra
\bea\label{eq:Rfluxalgebra}
[x^i,x^j] = \frac{\ii\ell_s^4}{3\hbar}\, R^{ijk}\, p_k \ , \qquad [x^i,p_j]=\ii\hbar\,\delta^i{}_j \qquad \mbox{and} \qquad [p_i,p_j] = 0 \ .
\eea
Note that \eqref{eq:Rfluxalgebra} for $d=3$ is formally the same as the magnetic monopole phase space algebra \eqref{eq:monopolealgebra} in the case of the linear magnetic field \eqref{eq:linearmagfield}, after applying the magnetic duality transformation of order four given by
\bea\label{eq:magneticduality}
x^i\longmapsto -p_i \ , \qquad p_i\longmapsto x^i \qquad \mbox{and} \qquad \mu_0\,\rho_m\,\hbar\, e\, \varepsilon_{ijk}\longmapsto -\frac{\ell_s^4}\hbar \, R^{ijk} \ ,
\eea
which in the absence of fluxes defines a canonical transformation. In particular, the locally non-geometric flux $R$ now appears as the curvature three-form of a two-form connection on a gerbe over momentum space~\cite{Mylonas2012}.
The quantization of the brackets \eqref{eq:Rfluxalgebra} can be captured by an explicit nonassociative star product on the phase space $T^*M$ of the original target space $M$~\cite{Mylonas2012,BL,Kupriyanov2015}, whose construction and physical implications will be discussed in detail later on.

\subsection{Caveats}

There are a few loopholes in the conformal field theory derivations
mentioned above that should be pointed out:
\begin{itemize}
\item[1.] The conformal field theory calculations are all performed in
  flat space with constant $H$-flux and constant dilaton. Such a
  background only satisfies the closed string equations of motion
  derived from the action \eqref{eq:Sgrav} to linear order in $H$ (in the critical dimension):
\eq\label{eq:SUGRAeqs}
\begin{aligned}
0&= {\rm Ric}_{ij}-\frac14\, H_i{}^{jk}\, H_{jkl} +2\,\nabla_i\nabla_j\Phi +
   O(\ell_s^4) \ , \\[4pt]
0&= -\frac12\, \nabla_kH^k{}_{ij}+\ell_s^2\, H_{ij}{}^k\,\nabla_k\Phi
   + O(\ell_s^4) \ , \\[4pt]
0&= \ell_s^2\, \Big((\nabla\Phi)^2-\frac12\, \nabla^2\Phi-\frac1{24}\,
   H^2 \Big) + O(\ell_s^4) \ .
\end{aligned}
\eqend
\item[2.] On a compact space with non-trivial topology such as the
  torus, the $H$-flux is quantized by virtue of generalized Dirac
  charge quantization in string theory. The true T-duality
  transformations connecting physically equivalent string theories are
  then valued in the discrete subgroup $O(d,d;\zed)\subset O(d,d)$. In this case the notion of
  linear order in $H$, as well as its extrapolation to non-geometric polarizations, is meaningless as $H$ cannot be continuously
  varied.
\item[3.] The triproduct violates the strong constraint between the
  background $R^{ijk}$ and the fluctuations around it~\cite{Blumenhagen:2013zpa}. This is not a
  serious problem if polarization can be achieved by a weaker
  constraint; this issue is currently under debate as the strong
  constraint indeed seems too stringent for some considerations. We elaborate on this point further in Section~\ref{sec:quantization}.
\item[4.] Nonassociativity does not appear in the algebra of conformal currents of the worldsheet theory, but rather through non-conformal fields such as $\CX^i$ above.
\item[5.] Thus far no construction is available of $H$-deformed
  \emph{graviton} vertex operators, whose correlation functions would clarify if and how a
  nonassociative deformation of the gravity theory defined by
  \eqref{eq:Sgrav} could be of relevance in closed string theory.
\end{itemize}
Despite these caveats, the geometric structures unveiled and their
novel physical implications are so rich and beautiful that work has
plugged along in this direction to further explore the ramifications
and relevance of
nonassociative geometry in closed string theory. 
For example, recent discussions reveal that Yang-Baxter deformations (in general a class of non-abelian T-duality) may be viewed as the open-closed string map \eqref{eq:openclosedrel} at the generalized supergravity level, see~\cite{Bakhmatov:2018apn} and references therein; in this setting the Jacobi identity yields the classical Yang-Baxter equation for the isometry group of the background. In the remainder of
this contribution we shall discuss further ways of naturally
understanding the origins of nonassociativity, and then proceed to
unravel some of the physics of this structure.

%%%%%%%%%%%%%%%%%%%%%%%%%%%%%%%%%%%%%%%%%%%%%%%%%%%%%%%
%%%%%%%%%%%%%%%%%%%%%%%%%%%%%%%%%%%%%%%%%%%%%%%%%%%%%%%

\section{Higher geometrization of non-geometry}
\label{sec:geometry}

In Section~\ref{sec:fluxes} we have discussed two geometric ways of making sense
of the globally and locally non-geometric frames of flux
compactifications: One through an extension of geometry into the realm
of generalized geometry and double field theory, and the other through
noncommutative and nonassociative deformations of the closed string
phase space. To better understand the relationship between these two points of view from a dynamical
perspective, we shall now show that the non-geometry of string
backgrounds is geometrized through a membrane sigma-model, suggesting
that the proper probes of these backgrounds should be open membranes
whose boundary modes are the closed string degrees of freedom. Such an approach was first suggested by~\cite{Halmagyi}, and developed in detail in~\cite{Mylonas2012}; see~\cite{Chatzistavrakidis2015,Bessho2015} for further developments. In
Section~\ref{sec:quantization} below we shall show that the
quantization of this sigma-model produces the nonassociative phase
space star product that we have been advertising.

To understand the idea behind this framework, let us recall how we treated the
open string sigma-model \eqref{eq:sigmamodel}. In the Seiberg-Witten scaling
limit it reduces to the action
\eqref{eq:sigmamodelscaling}, whose first order formalism describes a
two-dimensional topological field theory called the {`Poisson
  sigma-model'}. As we will discuss in Section~\ref{sec:quantization},
the quantization of this sigma-model reproduces the commutators
\eqref{eq:sigmamodelcomm} for a constant $B$-field, and more generally
quantizes the noncommutative
geometry defined by the Poisson bivector $\theta=B^{-1}$ on
$M=\real^d$ for vanishing
NS--NS flux $H=\dd B=0$.

At the other extreme, consider a constant non-vanishing NS--NS flux $H$ on
$M=\real^d$ and choose the symmetric gauge $B^\circ_{ij}=\frac13\, H_{ijk}\,
x^k$ for the Kalb-Ramond field. In this case the worldsheet action
\eqref{eq:sigmamodelscaling} can be written as
\bea\label{eq:WZterm}
S\big|_{g,\ell_s=0} = -\frac12\, \oint_{\Sigma_2}\, \frac13\,
H_{ijk}\, x^k\, \dd x^i\wedge\dd x^j =
-\frac12\, \int_{\Sigma_3}\, H_{ijk}\, \dd x^i\wedge \dd x^j\wedge
\dd x^k \ ,
\eea
where $\Sigma_3$ is a three-dimensional worldvolume with boundary
$\partial\Sigma_3=\Sigma_2$ and we used Stokes' theorem. This is just the well-known Wess-Zumino
action, which is needed in particular for a global formulation of the non-linear sigma-model when
the target space $M$ is e.g. a torus. The corresponding first order
formulation of this action is called the `$H$-twisted Poisson
sigma-model', which captures the topological dynamics of closed strings
in a non-trivial $H$-flux background. It is a particular case of a
three-dimensional topological field theory called a
`Courant sigma-model', which is defined on a \emph{membrane}
worldvolume.

The somewhat heuristic considerations above can be formalized within
the setting of generalized geometry, which provides a higher geometric
framework in which to study the geometric and non-geometric frames of closed
string theory as we have seen in Section~\ref{sec:fluxes} together with some aspects of its
extension to double field theory. 

\subsection{AKSZ theory}

The Alexandrov-Kontsevich-Schwarz-Zaboronsky (AKSZ) construction~\cite{Alexandrov:1995kv} is a geometric
framework for constructing Schwarz-type action functionals in the Batalin-Vilkovisky
formalism for $n{+}1$-dimensional topological sigma-models whose target space is a
symplectic Lie $n$-algebroid $E\to M$. These theories fit into a
geometric ladder describing $n{-}1$-dimensional degrees of freedom in background
fields, see e.g.~\cite{Ikedarev} for a review. The open string and membrane sigma-models alluded to above are the first two members of this hierarchy of theories, whose geometric structures we shall now discuss.

Consider first the case $n=1$. A Lie 1-algebroid is simply called a \emph{Lie algebroid} and consists of a vector bundle $E$ over the target space $M$ that sits in a diagram
\bea\label{eq:algebroiddiag}
\xymatrix{
E \ar[r]^\rho \ar[dr]_\pi & TM \ar[d] \\
 & M
}
\eea
where $\pi$ is the bundle projection and the anchor map $\rho$ to the
tangent bundle is compatible with a given Lie bracket $[s,s']_E$ on
sections $s,s'$ of $E$, in the sense that the following conditions
hold:
\begin{itemize}
\item \ \ $[s,f\,s']_E = f\,[s,s']_E + \big(\rho(s)f\big)\, s' \ , $
\item \ \ $\rho\big([s,s']_E\big) = \big[\rho(s),\rho(s')\big] \ , $
\end{itemize}
for any function $f\in C^\infty(M)$. The first axiom is the Leibniz rule, while the second axiom states that the anchor is a homomorphism between the Lie algebras of sections of $E$ and $TM$. If $M$ is a point, then a Lie
algebroid is the same thing as a Lie algebra $\mfg$ with zero anchor. On the other
hand, the tangent bundle $TM$ of any manifold $M$ is trivially a Lie
algebroid with the identity anchor map, and is called the
\emph{standard} Lie algebroid. The notion of a Lie algebroid thus generalizes these two simple examples simultaneously. For our purposes, the most relevant example of a Lie algebroid is the cotangent bundle $E=T^*M$ over a Poisson manifold $(M,\theta)$. In this case the anchor is $\rho=\theta^\sharp$ and the Lie bracket is the Koszul bracket
\bea
[\eta,\xi]_{\rm K} = \pounds_{\theta^\sharp\eta}\xi -
\pounds_{\theta^\sharp\xi}\eta - \dd\big(\theta(\eta,\xi)\big)
\eea
for one-forms $\eta,\xi$ on $M$; in particular, for functions $f,g$ one
has $[\dd f,\dd
g]_E=\dd\{f,g\}$ where $\{f,g\}=\theta(\dd f\wedge\dd g)$ is the Poisson
bracket induced by $\theta$.

The AKSZ sigma-model in this case is the Poisson sigma-model which is
described in more detail below. It quantizes point particles, viewed as
boundaries of open strings, in background magnetic fields, with underlying Hamiltonian dynamics governed by
the Poisson bracket. This leads to a noncommutative geometry.

Consider next the case $n=2$. A symplectic Lie 2-algebroid is the same thing as a \emph{Courant algebroid}, which is a vector bundle over $M$ sitting in a diagram like \eqref{eq:algebroiddiag} with a (not necessarily antisymmetric) bracket $[s,s']_E$ on sections $s,s'$ of $E$ and a fibrewise metric $\langle s,s'\rangle_E$ satisfying the following conditions:
\begin{itemize}
\item \ \ $[s,[s',s'']_E]_E=[[s,s']_E,s'']_E + [s',[s,s'']_E]_E \ , $
\item \ \ $[s,f\,s']_E = f\, [s,s']_E + \big(\rho(s)f\big)\,s' \ , $
\item \ \ $\rho\big([s,s']_E\big) = \big[\rho(s),\rho(s')\big] \ , $
\item \ \ $\rho(s'')\langle s,s'\rangle_E = \langle[s'',s]_E,s'\rangle_E + \langle s,[s'',s']_E\rangle_E \ . $
\end{itemize}
The first two properties endow the vector bundle $E$
with the structure of a `Leibniz algebroid', and are a generic feature
of all symplectic Lie $n$-algebroids; if the bracket is
antisymmetric, as in the case $n=1$, then the first axiom is
equivalent to the Jacobi identity. If $M$ is a point, then a Courant
algebroid is the same thing as a quadratic Lie algebra, i.e. a Lie
algebra $\mfg$ with an invariant inner product; a natural
class of examples is provided by the Drinfeld double $\mfg\oplus\mfg^*$ of a Lie bialgebra~$\mfg$. For our purposes, the significance of this higher
geometric structure is that if the generalized tangent bundle $E=TM\oplus
T^*M$ is endowed with the Dorfman bracket \eqref{eq:Dorfman}, the natural pairing
$\langle e_i,e^j\rangle=\delta_i{}^j$ between $TM$ and $T^*M$, and
anchor given by the projection $\rho(X+\xi)=X$, then $E$ forms a Courant algebroid called
the \emph{standard} Courant algebroid.

The AKSZ sigma-model in this case is the Courant sigma-model which is studied in detail below. It quantizes closed strings, regarded as boundaries of open membranes, in flux compactifications, with underlying worldsheet Hamiltonian dynamics governed by the Dorfman bracket. This leads to a nonassociative geometry.

The list continues, but the cases with $n\geqslant3$ are
not as well understood (see Section~\ref{sec:further} below). Let us now turn to the particular cases of AKSZ
sigma-models of direct relevance to us and clarify the statements made
above.

\subsection{Poisson sigma-models}

For $n=1$, the most general two-dimensional topological field theory
that can be constructed from the AKSZ theory is based on the
symplectic Lie algebroid $E=T^*M$, and gives rise to the \emph{Poisson
  sigma-model}~\cite{IkedaPoisson,SchallerStrobl} defined by the degree zero part of the AKSZ action
\bea\label{eq:Poissonsigmamodel}
S_{\rm AKSZ}^{(1)} = \int_{\Sigma_2}\, \Big( \xi_i\wedge\dd
x^i+\frac12\, \theta^{ij}(x)\, \xi_i\wedge\xi_j\Big) \ ,
\eea
where $x:\Sigma_2\to M$ are the string fields, $\xi$ are auxiliary
one-forms on $\Sigma_2$ valued in the cotangent bundle $T^*M$, and
$\theta=\frac12\, \theta^{ij}(x)\, \partial_i\wedge\partial_j$ is a
Poisson bivector on $M$. When $\theta=B^{-1}$ is non-degenerate,
integrating out the auxiliary one-form fields $\xi_i$ yields the topological
$B$-field amplitude \eqref{eq:sigmamodelscaling}. With suitable
Dirichlet boundary conditions on the fields, the perturbative expansion of the
corresponding path integral leads to the Kontsevich formality
maps~\cite{Kontsevich:1997vb,Cattaneo:1999fm}, which we will discuss in
Section~\ref{sec:quantization} below. The on-shell condition derived from this
action is equivalent to $[\theta,\theta]_{\rm S}=0$, i.e. that the bivector
$\theta$ defines a Poisson structure on $M$, but the sigma-model 
makes sense off-shell as well, for instance when $\theta$ is a twisted
Poisson structure.

\subsection{Courant sigma-models}

For $n=2$, we take the symplectic Lie 2-algebroid to be the standard
Courant algebroid $E=TM\oplus T^*M$, and define
the three-dimensional topological field theory on a membrane
worldvolume $\Sigma_3$ by the degree zero part of the AKSZ action
\bea
S_{\rm AKSZ}^{(2)} = \int_{\Sigma_3}\, \Big(\phi_i\wedge \dd x^i
+\frac12\, \eta_{IJ}\, \alpha^I\wedge\dd \alpha^J-\rho_I{}^i\,
\phi_i\wedge\alpha^I + \frac16\, T_{IJK}(x)\,
\alpha^I\wedge\alpha^J\wedge\alpha^K\Big) \ ,
\eea
where $x:\Sigma_3\to M$ are the membrane fields, $\alpha$ are
one-forms on $\Sigma_3$ valued in $E$, and $\phi$ are auxiliary
two-forms on $\Sigma_3$ valued in the cotangent bundle $T^*M$. The fibre
metric has components $\eta_{IJ}=\langle s_I,s_J\rangle_E$ in a local
basis of sections $s_I$ of $E$, the anchor map has components
$\rho(s_I) = \rho_I{}^i\, e_i$, and the three-form $T_{IJK}(x) = \langle
s_I,[s_J,s_K]_E\rangle_E$ can accomodate all four geometric and
non-geometric fluxes in the T-duality orbit \eqref{eq:Tdualitychain}. This is called the \emph{Courant sigma-model}~\cite{RoytenbergAKSZ}.

Let us begin by describing the geometric $H$-flux frame. We write the
one-forms as $\alpha=(\alpha^I)=(\alpha^i,\xi_i)$ corresponding to the
splitting $E=TM\oplus T^*M$ and use the $H$-twisted Dorfman bracket
from \eqref{eq:HtwistedDorfman} with vanishing torsion to write
the open membrane action
\bea
S_H^{(2)} = \int_{\Sigma_3}\, \Big(\phi_i\wedge \dd x^i +
\alpha^i\wedge\dd\xi_i - \phi_i\wedge\alpha^i + \frac16\, H_{ijk}(x)\,
\alpha^i\wedge\alpha^j\wedge\alpha^k \Big) \ .
\eea
With suitable Dirichlet boundary conditions on the fields, by
integrating out the auxiliary two-form fields $\phi_i$ we arrive at the action
on the closed string worldsheet $\Sigma_2=\partial\Sigma_3$ given by 
\bea
S_H^{(2)} = \oint_{\Sigma_2}\, \xi_i\wedge\dd x^i + \int_{\Sigma_3}\,
\frac16\, H_{ijk}(x)\, \dd x^i\wedge\dd x^j\wedge\dd x^k \ .
\eea
This is just the Wess-Zumino action, which for constant $H$-flux falls
entirely on the worldsheet $\Sigma_2$ as in \eqref{eq:WZterm}. One can
also add a boundary perturbation to the membrane sigma-model by an
arbitrary bivector $\theta$ on $M$ of the form given in
\eqref{eq:Poissonsigmamodel}. This defines the action of the
\emph{$H$-twisted Poisson sigma-model}~\cite{AlekseevStroblWZW}, whose on-shell conditions imply
$[\theta,\theta]_{\rm S}=\bigwedge^3\theta^\sharp H$, so that $\theta$
defines an $H$-twisted Poisson structure on the target space $M$; in
particular, the Jacobi identity for the corresponding bracket defined
by $\theta$ is violated by the NS--NS three-form flux.

Now let us consider the locally non-geometric $R$-flux frame. Using the
$R$-twisted Dorfman bracket from \eqref{eq:RtwistedDorfman} with
vanishing torsion and $Q$-flux, and with
notation as above, after integrating out the auxiliary fields $\phi_i$
again we arrive at the open membrane sigma-model with (rescaled) $R$-flux given
by the action
\bea
S_R^{(2)} = \int_{\Sigma_3}\, \Big(\dd\xi_i\wedge\dd x^i + \frac{\ell_s^4}{18\hbar^2} \,
R^{ijk}(x)\, \xi_i\wedge\xi_j\wedge\xi_k\Big) \ .
\eea
When the $R$-flux is constant, with suitable Dirichlet boundary
conditions the equations of motion for the
boundary string fields $x^i$ imply $\xi_i = \dd p_i$ for local fields $p$ valued
in the fibres of the cotangent bundle $T^*M$ (up to harmonic forms on $\Sigma_2$),
so that using Stokes' theorem and linearizing the resulting action
with auxiliary one-form fields $\eta_I$ valued in $E$ leads to the boundary action~\cite{Mylonas2012}
\bea\label{eq:PoissonRflux}
S_R^{(2)} = \oint_{\Sigma_2}\, \Big(\eta_I\wedge \dd x^I + \frac12\,
\Theta^{IJ} \, \eta_I\wedge\eta_J\Big) \ .
\eea
Here $x=(x^I)=(x^i,p_i):\Sigma_2\to T^*M$ are string fields valued
in the cotangent bundle of $M$, so that the effective target space is
now the \emph{phase space}. This action defines a Poisson sigma-model
for the bivector
\bea\label{eq:twistedRflux}
\Theta = (\Theta^{IJ}) = \bigg( \begin{matrix} R(p) & 1_d \\ -1_d &
  0 \end{matrix} \bigg) \qquad \mbox{with} \quad R(p)^{ij} = \frac{\ell_s^4}{3\hbar^2} \, R^{ijk}\,
p_k \ .
\eea
The corresponding quantum phase space brackets
\bea\label{eq:Rcommutators}
[x^I,x^J] = \ii\hbar\, \Theta^{IJ}(x)
\eea
coincide precisely with the $R$-space commutation relations
\eqref{eq:Rfluxalgebra}. Note that for $d=3$ this twisted Poisson structure is
formally identical to \eqref{eq:monopoleNC} with a linear
magnetic field \eqref{eq:linearmagfield} under the magnetic duality
transformation \eqref{eq:magneticduality}, and in particular the
commutators \eqref{eq:Rcommutators} together with the
corresponding Jacobiators
\bea
[x^I,x^J,x^K] = -\hbar^2\,[\Theta,\Theta]_{\rm S}^{IJK} =
\bigg( \begin{matrix} \ell_s^4\, R^{ijk} & 0 \\ 0 & 0 \end{matrix} \bigg)
\eea
yield a noncommutative and nonassociative phase space geometry.

One important caveat with this derivation is that the setting of the
$Q$-flux to zero in \eqref{eq:RtwistedDorfman} does not define a
Courant algebroid structure, as is evident from
\eqref{eq:QRtheta}. This simply reflects the fact that the
nonassociative geometry of the $R$-flux frame violates the strong constraint
of double field theory, as mentioned previously, so that the
corresponding membrane sigma-model does not define a Courant
sigma-model. It can, however, be obtained in a precise way via
projection from a proper Courant sigma-model defined on the doubled
space of double field theory that incorporates all fluxes of
\eqref{eq:Tdualitychain} in a manifestly T-duality invariant way~\cite{Chatzistavrakidis:2018ztm}, which clarifies precisely the
geometric algebroid structure underlying the non-geometric frames of
closed string theory, and also how gauge invariance is restored in the
non-geometric membrane sigma-models through the Bianchi identities
among the fluxes of the T-duality chain
\eqref{eq:Tdualitychain}. In~\cite{Chatzistavrakidis:2018ztm} it is
also shown how to treat the noncommutative geometry of the globally non-geometric $Q$-flux frame in
an analogous manner, and how to generally treat the non-geometry through Courant
sigma-models via the open-closed string reparameterization
\eqref{eq:openclosedrel} of the algebroid structure maps, thereby clarifying the evident similarities we have seen between the phase space and double field theory frameworks for non-geometry (see also \cite{Freidel:2015pka,Freidel:2017yuv}).

%%%%%%%%%%%%%%%%%%%%%%%%%%%%%%%%%%%%%%%%%%%%%%%%%%%%%%%
%%%%%%%%%%%%%%%%%%%%%%%%%%%%%%%%%%%%%%%%%%%%%%%%%%%%%%%

\section{Quantization of non-geometric backgrounds}
\label{sec:quantization}

In this section we will describe the quantization of the twisted Poisson structure \eqref{eq:twistedRflux} describing the nonassociative geometry of the closed string phase space in the $R$-flux frame, and then present some of its far reaching applications to non-geometric string theory.

\subsection{Quantization of topological string theory}

Suitable functional integrals in the $R$-flux Poisson sigma-model \eqref{eq:PoissonRflux} reproduce Kontsevich's formality maps for global deformation quantization of twisted Poisson manifolds~\cite{Kontsevich:1997vb}. The formality maps $U_n$ take $n$ multivector fields $\CX_1,\dots,\CX_n$ of degrees $k_1,\dots,k_n$ to multidifferential operators $D_\Gamma(\CX_1,\dots,\CX_n)$ of degree $2-2n+k_1+\cdots+k_n$. They have a combinatorial expansion as a sum over graphs
\bea\label{eq:Unsum}
U_n(\CX_1,\dots,\CX_n) = \sum_{\Gamma_n\in G_n} \, w_{\Gamma_n} \, D_{\Gamma_n}(\CX_1,\dots,\CX_n) \ ,
\eea 
where $G_n$ is the set of admissible graphs $\Gamma_n$, with $n$
vertices and edges $\big\{e_i^1,\dots,e_i^{k_i}\big\}_{i=1}^n$, which can be
drawn in the configuration space $\bbH_n$ of $n$ points on the
hyperbolic plane with prescribed weights
\bea\label{eq:weights}
w_{\Gamma_n} = \frac1{(2\pi)^{\sum_i\,k_i}}\, \int_{\bbH_n} \ \bigwedge_{i=1}^n\, \big(\dd\phi_{e_i^1}\wedge\cdots\wedge\dd\phi_{e_i^{k_i}}\big) \ .
\eea

In particular, inserting any bivector $\CX_i=\Theta=\frac12\, \Theta^{IJ}\, \partial_I\wedge \partial_J$ into all slots gives a sum of bi{-}differential operators, which can be represented diagrammatically by graphs with edges emanating along the legs of $\Theta$ and operating on functions $f,g$ sitting on the boundary of the hyperbolic plane. This defines the star product
\bea\label{eq:fstargUn}
f\star g = \sum_{n=0}^\infty\, \frac{(\ii\hbar)^n}{n!}\, U_n(\Theta,\dots,\Theta)(f,g) =: \CCP(\Theta)(f,g) \ ,
\eea
which up to order $\hbar^2$ is given explicitly by
\bea
\begin{aligned}
f\star g &= f\cdot g + \frac{\ii\hbar}2\, \Theta^{IJ}\, \partial_If\cdot\partial_Jg - \frac{\hbar^2}4\, \Theta^{IJ}\, \Theta^{KL}\, \partial_I\partial_Kf\cdot \partial_J\partial_Lg \\
& \qquad\qquad-\frac{\hbar^2}6\, \Theta^{IJ}\, \partial_J\Theta^{KL}\, \big(\partial_I\partial_Kf\cdot \partial_Lg - \partial_Kf\cdot \partial_I\partial_Lg\big)+O(\hbar^3) \ .
\end{aligned}
\eea
The order $\hbar$ term is the semiclassical contribution which is proportional to the classical bracket $\{f,g\}=\Theta(\dd f\wedge\dd g)$ defined by the bivector $\Theta$.

As a simple example, let us consider a constant bivector $\theta$ in the Poisson sigma-model \eqref{eq:Poissonsigmamodel}. The basic graph $\Gamma_1$ with a single vertex and two edges contributes the weight
\bea
w_{\Gamma_1} = \frac1{(2\pi)^2}\, \int_0^{2\pi}\, \dd\psi \ \int_0^{\psi}\, \dd\phi = \frac1{(2\pi)^2}\, \Big[\frac12\, \psi^2\Big]_0^{2\pi} = \frac12 \ .
\eea
In this case all graphs and hence weight integrals \eqref{eq:weights} factorize in terms of this basic graph $\Gamma_1$, so that the sum \eqref{eq:Unsum} truncates to
\bea
U_n(\theta,\dots,\theta) = \Big(\frac12\Big)^n\, \theta^{i_1j_1}\cdots\theta^{i_nj_n} \, \big(\partial_{i_1}\cdots\partial_{i_n} \big) \otimes \big(\partial_{j_1}\cdots\partial_{j_n} \big) \ .
\eea
The star product \eqref{eq:fstargUn} thus becomes
\bea
f\star g = \sum_{n=0}^\infty\, \frac{(\ii\hbar)^n}{n!}\, \Big(\frac12\Big)^n\, \theta^{i_1j_1}\cdots\theta^{i_nj_n}\, \partial_{i_1}\cdots\partial_{i_n}f \cdot \partial_{j_1}\cdots\partial_{j_n}g \ ,
\eea
which is just the Moyal-Weyl star product \eqref{eq:Moyal}.

In general, if $\Theta$ is not constant but still satisfies $\Theta^{IJ}\,\partial_J\Theta^{KL}=0$, then the series \eqref{eq:fstargUn} again exponentiates exactly as in the case of a constant bivector which led to the Moyal-Weyl star product. This is the case for the bivector \eqref{eq:twistedRflux} with constant $R$-flux, leading to the star product~\cite{Mylonas2012}
\bea\label{eq:NAstarproduct}
f\star g = \cdot\, \exp\Big(\frac{\ii\hbar}2\,\Big[\frac{\ell_s^4}{3\hbar^2}\, R^{ijk}\, p_k\, \partial_i\otimes\partial_j+\partial_i\otimes \tilde\partial{}^i-\tilde\partial{}^i\otimes \partial_i\Big]\Big)(f\otimes g) \ ,
\eea
where $\tilde\partial{}^i=\frac\partial{\partial p_i}$ denote momentum
derivatives; in Fourier space it reads as
\bea
(f\star g)(x,p)=\int\, \dd k\ \dd\tilde k \ \int\, \dd q\ \dd\tilde q \ \tilde f(k,\tilde k)\, \tilde g(q,\tilde q) \, \e^{\frac{\ii\hbar}2\, (\tilde k{}^i \, q_i-k_i \, \tilde q{}^i)} \, \e^{-\frac{\ii\ell_s^4}{6\hbar}\, R^{ijk} \, k_i\, q_j\, p_k} \, \e^{\ii(k+q)_i\, x^i} \ .
\eea
This is a deformation by $R$-flux of the usual phase
space Moyal product for deformation quantization in ordinary quantum
mechanics, see e.g.~\cite{Zachos2001} for a concise review. Other approaches to deriving this star product can be found in~\cite{Mylonas2012,BL,Mylonas2013,Kupriyanov2015}. Various formal properties of this and other classes of nonassociative star products in deformation quantization are discussed in~\cite{Bojowald:2016lnl,Vassilevich:2018gkl}.

The formality maps $U_n$ define quasi-isomorphisms between 
differential graded $L_\infty$-algebras, relating Schouten brackets $[
\ , \ ]_{\rm S}$,
i.e. the obvious extensions of the Lie bracket of vector fields to
multivector fields, to Gerstenhaber brackets $[ \ , \ ]_{\rm G}$, i.e. the obvious
extensions of the commutator of differential operators to
multidifferential operators. As such they satisfy a set of `formality
conditions'. In particular, one of these conditions reads as
\bea
\ii\hbar\, \CCP([\Theta,\Theta]_{\rm S}) = [\CCP(\Theta),\star]_{\rm
  G} \ .
\eea
This quantifies nonassociativity, since when applied to triples of
functions $f,g,h$, it follows that $[\Theta,\Theta]_{\rm S}\neq0$ if and
only if
\bea
(f\star g)\star h - f\star(g\star h)\neq0 \ .
\eea
That is, $\Theta$ is a Poisson bivector if and only if the star
product is associative. For the $R$-space with constant locally
non-geometric flux, the star commutators
\bea\label{eq:starcomm}
[x^I,x^J]_\star := x^I\star x^J - x^J\star x^I =
\ii\hbar\, \Theta^{IJ}(x) 
\eea
have the corresponding non-vanishing star Jacobiators
\bea\label{eq:starJac}
[x^i,x^j,x^k]_\star = \ell_s^4\, R^{ijk} \ .
\eea
For general phase space functions, the nonassociativity of the star product can
be expressed in closed form as
\bea\label{eq:associator}
(f\star g)\star h = \varphi_{f,g,h}\big(f\star(g\star h)\big) := \star\, \exp\Big(\frac{\ell_s^4}6\,
R^{ijk}\, \partial_i\otimes \partial_j\otimes \partial_k
\Big)\big(f\otimes(g\otimes h)\big) \ ,
\eea
where the associators
\bea
{f}\star( {g}\star
 {h}) \ \xrightarrow{ \
  \varphi_{{f},{g}, {h}} \ } \
({f}\star {g})\star {h} 
\eea
satisfy pentagon relations implied by the (higher) formality conditions~\cite{Mylonas2012}, which state that the diagram
\bea\label{eq:pentagon}
\xymatrix{
 & ( {f} \star {g} )\star(
 {h} \star {k} ) \ar[ddl]_{\varphi_{{f} \star {g},
     {h}, {k}\ }}
 & \\
 & & \\
\big(( {f} \star {g} )\star
{h} \big)\star {k}
   & & {f} \star \big( {g} \star
( {h} \star {k} ) \big) \ar[uul]_{\varphi_{{f},{g},
     {h}\star {k}}} \ar[dd]^{1 \otimes
   \varphi_{{g},
     {h}, {k}}}\\
 & & \\
\big( {f} \star ( {g} \star
{h}) \big)\star {k} \ 
\ar[uu]^{\varphi_{{f},{g}, 
     {h}} \otimes 1} & & \ {f} \star
   \big((  {g} \star
{h}) \star {k}  \big) \ar[ll]^{\varphi_{{f},{g} \star 
     {h}, {k}}}
}
\eea
commutes for all functions $f,g,h,k$. In particular, the translation group three-cocycle discussed in Section~\ref{sec:monopole} is realised here through a nonassociative version of the Baker-Campbell-Hausdorff formula which yields
\bea
\big(\e^{\ii k_I\, x^I}\star \e^{\ii q_I\, x^I}\big)\star \e^{\ii r_I\, x^I} = \e^{\frac{\ii\ell_s^4}{6}\, R^{ijk} \, k_i\, q_j\, r_k} \ \e^{\ii k_I\, x^I}\star \big(\e^{\ii q_I\, x^I}\star \e^{\ii r_I\, x^I}\big) \ .
\eea

In the remainder of this section we consider various applications of
this nonassociative phase space star product formalism.

\subsection{Triproducts}

We shall now address Question (Q2) from Section~\ref{sec:intro}. Given
$n$ functions $f_1,\dots,f_n\in C^\infty(M)$ on the target space $M$, using
the nonassociative phase space star product we define the
\emph{triproducts}
\bea\label{eq:tristarproducts}
f_1\tri f_2\tri \cdots \tri f_n := \lim_{p\to0} \ f_1\star \big(f_2\star \big(\cdots
(f_{n-1}\star f_n)\cdots \big)\big) \ ,
\eea
where we have chosen a particular bracketing for the star products of
functions. It was shown by~\cite{Aschieri:2015roa} that the result of
this operation can be expressed in terms of a tridifferential operator
as
\bea
f_1\tri f_2\tri \cdots \tri f_n = \cdot\,
\exp\Big(-\frac{\ell_s^4}{6}\, \sum_{1\leqslant a<b<c\leqslant n}\,
R^{ijk}\, \partial^a_i\otimes\partial^b_j\otimes\partial^c_k\Big) (f_1\otimes
f_2\otimes \cdots \otimes f_n) \ ,
\eea
where $\partial^a_i$ denotes the action of the derivative $\partial_i$
in the $a$-th factor of the tensor product $f_1\otimes
f_2\otimes\cdots\otimes f_n$.
This remarkable algebraic structure has the following 
properties. 

Firstly, it agrees with the triproducts of tachyon vertex
operators after T-duality in worldsheet perturbation theory around
flat space with constant $H$-flux, as we discussed earlier. This
formula was derived to linear order in the $R$-flux in~\cite{Blumenhagen2011}, where its all
orders exponential form was conjectured, and
subsequently extrapolated to non-constant fluxes appropriate to curved
backgrounds $M$ in~\cite{Blumenhagen:2013zpa}. Here we find an
explicit calculational \emph{derivation} of these conjectural
expressions from the nonassociative phase space star product, which
lends further credibility to the purported description of the locally
non-geometric flux background.

Secondly, since the momentum is set to $p=0$, for $n=2$ there is no
deformation of the usual pointwise product of fields: $f\tri g =
f\cdot g$, as anticipated from the closed string theory considerations
of Section~\ref{sec:intro}. For $n=3$ the
triproduct was originally introduced in~\cite{Takhtajan:1993vr} as a
proposal for the quantization of Nambu brackets~\cite{Nambu}, which we will discuss in
more detail later on. It quantizes the
tribracket defined by \eqref{eq:tribracketprod}, and in particular one has
\bea\label{eq:triprodbracket}
[x^i,x^j,x^k]_{\tri} = \ell_s^4\, R^{ijk} \ ,
\eea
which agrees with the Jacobiator \eqref{eq:starJac}.

Thirdly, the triproducts exhibit on-shell associativity:
\bea
\int\, \dd x \ f_1\tri \cdots \tri f_n = \int\, \dd x \ f_1 \cdots f_n
\ ,
\eea
which follows from on-shell momentum
conservation~\cite{Aschieri:2015roa}. As we discussed earlier, this is
expected and in fact \emph{necessary} from the string theory
perspective: As on-shell closed string theory is described a
two-dimensional quantum field theory, it is described by an
associative operator algebra. In particular, the on-shell
associativity agrees with the crossing symmetry of $n$-point
correlation functions on the sphere $S^2$.

Finally, as also mentioned before, the triproducts violate the strong
constraint of double field theory. Recalling the microscopic origin of
the $R$-flux in terms of a bivector from \eqref{eq:Rfluxtheta}, the
expansion of the triproduct for $n=3$ reads as
\bea
f\tri g\tri h =f\cdot g\cdot h - \frac{\ell_s^4}{6}\,
\tilde\partial{}^{[i}\tilde\theta^{jk]}\, \partial_if\cdot \partial_jg\cdot \partial_kh+
O(\ell_s^8) \ ,
\eea
and the corrections to the pointwise product of fields vanishes by the
strong constraint. Thus the nonassociative geometry of the closed
string background is only probed if the strong constraint between the
background and fluctuations is weakened, a point which is presently
under debate.

\subsection{Nonassociative quantum mechanics}

Let us now investigate some of the physical consequences of the
nonassociative deformation of quantum mechanics in the $R$-flux background. The standard
operator-state formulation of quantum mechanics cannot handle
nonassociative structures: Operators which act on a Hilbert
space necessarily associate. However, we can generalize the phase space
formulation of quantum mechanics~\cite{Zachos2001} to provide a completely quantitative
and physically viable formulation of
nonassociative quantum mechanics~\cite{Mylonas2013}. Other approaches to nonassociative quantum mechanics based on the magnetic monopole algebra \eqref{eq:monopolealgebra} are found in~\cite{Bojowald:2014oea,Bojowald:2015cha}.

The idea behind phase space quantum mechanics is to treat position and
momentum variables on equal footing. In this setting generic
`operators' become complex-valued functions on phase space, with the
operator product provided by the star product and traces given by
integration of functions; an `observable' is then a real-valued
function on phase space. Dynamics of observables $A$ are governed by
Heisenberg-type time evolution equations
\bea
\dot A = \frac\ii\hbar\, [H,A]_\star \ ,
\eea
for a given classical Hamiltonian $H$. 

The key properties of the
nonassociative phase space star product \eqref{eq:NAstarproduct} that are needed in this
description are as follows. Firstly, since the star product $f\star g$
again differs from $f\cdot g$ by a total derivative, and likewise
$f\star(g\star h)$ differs from $(f\star g)\star h$ by a total
derivative, one again has the 2-cyclicity and 3-cyclicity properties
\eqref{eq:2cyclic} and \eqref{eq:3cyclic} that we encountered for the
Moyal-Weyl product. Recall that these features were crucial for
compatibility of the star product formalism with the axioms of
conformal field theory; in the present context, they refer to the fact
that the star product leads to a traceless commutator and
associator. However, there are in general inequivalent quartic
expressions, for example
\bea
\int\, \dd x\ f\star\big(g\star(h\star k)\big) \neq \int\, \dd x\
f\star\big((g\star h)\star k\big) \ ,
\eea
see~\cite{Mylonas2013,Mylonas:2014kua} for a detailed discussion. Using 2-cyclicity we deduce the crucial positivity
property
\bea
\int\, \dd x\ \overline{f}\star f \geqslant 0 \ .
\eea
Finally, the nonassociative star product is Hermitian, in the sense
that $\overline{f\star g} = \overline{g}\star\overline{f}$, and
unital, in the sense that the constant unit function $1$ is still an
identity element for the nonassociative star product algebra: $f\star
1 = f = 1 \star f$. All of these properties are completely analogous
to those of the usual Moyal product in canonical phase space quantum
mechanics and they mimick the expected features of the operator
product; indeed, the star product \eqref{eq:NAstarproduct} for the
constant $R$-flux background is the simplest example of a
nonassociative star product and provides the natural extension of
the Moyal-Weyl star product \eqref{eq:Moyal} to the non-geometric
string background.

A `state' in nonassociative quantum mechanics is determined by a set
of square-integrable phase space wavefunctions $\psi_a$ which are
normalized:
\bea
\int\, \dd x\ |\psi_a|^2 = 1 \ ,
\eea
and a collection of statistical probabilities $\mu_a\in[0,1]$ obeying
\bea
\sum_a\, \mu_a = 1 \ .
\eea
The `expectation value' of an operator $A$ is then defined by
\bea\label{eq:expectation}
\langle A\rangle_\psi = \sum_a\, \mu_a \ \int\, \dd x\
\overline{\psi_a}\star (A\star \psi_a) = \int\, \dd x\ W_\psi\cdot A
\ ,
\eea
where we introduced the state function
\bea
W_\psi = \sum_a\, \mu_a\, \psi_a\star\overline{\psi_a} 
\eea
which is the analogue of the Wigner distribution function; in particular,
it is real-valued and normalized: $\langle1\rangle_\psi = \int\, \dd
x\ W_\psi = 1$. However, even in canonical phase space quantum
mechanics~\cite{Zachos2001}, one of the pitfalls of the formalism is that state
functions are not necessarily non-negative and so only determine
quasi-probability distribution functions in general.

As a simple example to familiarize ourselves with the phase space
formulation of quantum mechanics, let us consider the simplest case of
the free particle with Hamiltonian
\bea
H(x,p)=\sum_{i=1}^d\, \frac{p_i^2}{2m}
\eea
and vanishing $R$-flux. We denote the corresponding Moyal star product
from \eqref{eq:NAstarproduct} with $R^{ijk}=0$  by
$\star_0$ and look for solutions of the time-independent Schr\"odinger
equation
\bea
H\star_0 W_\psi = E\, W_\psi \ .
\eea
This can be expressed as a second order partial differential equation
\bea
\frac1{2m}\, \sum_{i=1}^d\, \Big( p_i^2+\ii\hbar\, \partial_i -
\frac{\hbar^2}4\, \partial_i^2\Big) W_\psi = E\, W_\psi
\eea
for the real-valued Wigner distribution function $W_\psi(x,p)$. It collapses to a pair of partial differential equations, its real and imaginary parts. The imaginary part
\bea
\sum_{i=1}^d\, p_i\, \partial_i W_\psi(x,p)=0
\eea
restricts $W_\psi(x,p)=W_\psi(p)$ to be independent of $x$. The real part
\bea
\sum_{i=1}^d\, \Big(p_i^2-\frac{\hbar^2}4 \, \partial_i^2-2m\, E \Big)
W_\psi(p) = 0
\eea
is satisfied for arbitrary real functions $W_\psi(p)$ of momentum with the energy eigenvalues
\bea
E=E_p= \sum_{i=1}^d\, \frac{p_i^2}{2m} \ .
\eea
This calculation can be extended to include an anisotropic harmonic
oscillator potential of frequencies $\omega_i$, for which the real part of
the second order partial differential equation can be separated into
Laguerre equations in the phase space variables $z_i=\frac2{m\,\hbar}\, (m\,
\omega_i^2\, x_i^2+p_i^2)$ for $i=1,\dots,d$~\cite{Zachos2001}. The corresponding solutions in terms of Laguerre
polynomials reproduce the anticipated harmonic oscillator spectrum 
\bea
E = E_n = \sum_{i=1}^d\, \hbar\,\omega_i\, \Big(n_i+\frac12\Big) \qquad
\mbox{with} \quad n=(n_i)\in\nat^d_0
\eea
in
$d$ dimensions. 
The extension of these considerations to a free particle moving in a
non-vanishing $R$-flux background is discussed
in~\cite{Szabo:2017yxd}, and from the alternative associative
framework of symplectic realisation of the twisted Poisson structure
\eqref{eq:twistedRflux} in~\cite{Kupriyanov:2018xji}. 

The operator-state correspondence can be described in this framework
as follows. The idea is that operators should still be combined
together with an associative operation, since nonassociativity should
not affect correlation functions. To this end we introduce two
conjugate ``operator algebras'' by defining left and right
compositions of observables $A,B$ as
\bea
(A\circ B)\star f := A\star(B\star f) \qquad \mbox{and} \qquad
f\star(A\,\bar\circ\, B) := (f\star A)\star B \ ,
\eea
for a test function $f$.
By definition the composition products are associative, since
\bea
(A_1\circ A_2\circ \cdots \circ A_n)\star f = A_1\star\big(A_2\star
\cdots \star(A_n\star f)\cdots \big) \ .
\eea
They are furthermore unital: $A\circ 1 = A = 1\circ A$, and the
composition products of basic coordinate monomials coincide with their
star products, e.g. $x^i\circ x^i = x^i\star x^i = (x^i)^2$ and
$p_i\circ p_i = p_i\star p_i = (p_i)^2$. However, in general the
composition product $A\circ B$ of two functions $A$ and $B$ is not a
function, but rather a differential operator; the composition products
arise as the embedding of the nonassociative star product algebra of
functions as a \emph{subspace} (but not as a subalgebra) of an
associative algebra of differential operators on phase space. In the
conventional frameworks where $\star$ is associative, the composition
product and the star product coincide. The physical and geometric meaning of the associative composition algebra of differential operators is elucidated by~\cite{Kupriyanov:2018xji} from the perspective of symplectic realisation of the twisted Poisson structure \eqref{eq:twistedRflux}.

We make the convention that the conjugate composition product $\bar\circ$
is always evaluated before $\circ$. Then a `state' $\rho_\psi$ is an
expression of the form
\bea
\rho_\psi = \sum_a\, \mu_a\, \psi_a\,\bar\circ\,\overline{\psi_a} \ .
\eea
This quantity should be thought of as a special representation of a ``density matrix'', which in
the associative setting would be the same as the Wigner distribution
function; in particular, the expectation values \eqref{eq:expectation}
of operators are
determined via traces with $\rho_\psi$ through
\bea
\langle A\rangle_\psi = \int\, \dd x\ A\star \rho_\psi \ .
\eea
The expectation values of compositions of operators are then given by
\bea
\begin{aligned}
\langle A_1\circ A_2\circ \cdots \circ A_n\rangle_\psi &= \int\, \dd x\
(A_1\circ A_2\circ \cdots \circ A_n)\star \rho_\psi \\[4pt] 
&= \sum_a\, \mu_a \
\int\, \dd x\ \big(A_1\star(A_2\star\cdots\star(A_n\star
\psi_a)\cdots)\big)\star \overline{\psi_a} \ .
\end{aligned}
\eea

Let us make some simple consistency checks of this purported
nonassociative version of quantum mechanics. First, let us check
reality. A straightforward calculation using the above definitions and
the properties of
the nonassociative star product gives
\bea
\overline{\langle A\rangle_\psi} = \sum_a\, \mu_a \ \int\, \dd x\
\overline{(A\star \psi_a)}\star\psi_a = \sum_a\, \mu_a \ \int\, \dd x\
\overline{\psi_a}\star (\, \overline{A}\star\psi_a) = \langle\,
\overline{A}\, \rangle_\psi \ .
\eea
It follows that the expectation values of observables, i.e. real
functions, are real. Similarly, one checks positivity: $\langle\,
\overline{A}\circ A\rangle_\psi\geqslant 0$. These derivations are
carried out in complete analogy with the corresponding calculations in
canonical phase space quantum mechanics; however, in the
nonassociative case an extra line or two is always required in the
computation.

Next we check that observables, i.e. real functions $A=\overline{A}$,
have real eigenvalues. For this, consider the ``star-genvalue
equation'' $A\star f = \lambda\, f$ for a complex number $\lambda$;
complex conjugation of this equation using Hermiticity of the star
product yields $\overline{f}\star\overline{A} = \overline{\lambda}\
\overline{f}$. From this we calculate
\bea
\overline{f}\star(A\star f) - (\,\overline{f}\star A)\star f =
(\lambda-\overline{\lambda}\,) (\,\overline{f}\star f) \ .
\eea
In the associative case this would immediately imply that
$\lambda=\overline{\lambda}$ is real, since the left-hand side would
automatically vanish. In the nonassociative case this is not generally
true, but using 3-cyclicity and 2-cyclicity we can integrate both
sides this equation to get
\bea
0 = (\lambda-\overline{\lambda}\,)\, \int\, \dd x\ |f|^2 \ .
\eea
Since the integral on the right-hand side is non-zero for $f\neq0$, it
follows that $\lambda = \overline{\lambda}$. A similar calculation
establishes that eigenfunctions with different eigenvalues are
orthogonal in the $L^2$-inner product.

\subsection{Spacetime quantization}

As a concrete application of the formalism, we will now show how
nonassociative quantum mechanics in the $R$-flux background leads to a
coarse-graining of spacetime. There are several ways in which to see
this. In standard quantum mechanics it is a fundamental result that
pairs of noncommuting operators cannot have simultaneous
eigenvalues. Similarly, in the present case we can show that triples
of nonassociating operators cannot have simultaneous
eigenvalues. Suppose that $W_\psi$ is a state function which simultaneously
diagonalizes a triple of basic phase space coordinate operators: $x^I\star
W_\psi=\lambda^I\, W_\psi$, $x^J\star
W_\psi=\lambda^J\, W_\psi$ and $x^K\star
W_\psi=\lambda^K\, W_\psi$ for eigenvalues $\lambda^I, \lambda^J,
\lambda^K\in\real$. By repeatedly applying the cyclicity properties of
the star product we compute
\bea
\begin{aligned}
\int\, \dd x\ \big((x^I\star x^J)\star x^K\big)\star W_\psi &= \int\,
\dd x\ (x^I\star x^J)\star(x^K\star W_\psi) \\[4pt]
&= \lambda^K\, \int\, \dd x\ (x^I\star x^J)\star W_\psi \\[4pt]
&= \lambda^K\, \int\, \dd x\ x^I\star(x^J\star W_\psi) \\[4pt]
&= \lambda^K\, \lambda^J\, \lambda^I \ ,
\end{aligned}
\eea
and similarly
\bea
\int\, \dd x\ \big(x^I\star(x^J\star x^K)\big)\star W_\psi =
\lambda^I\, \lambda^K\, \lambda^J \ .
\eea
Taking the difference of these two equations implies the vanishing Jacobiators $[x^I,x^J,x^K]_\star =
0$, which contradicts the basic nonassociative deformation provided by the
position coordinate operators in \eqref{eq:starJac}. This implies a
coarse-graining of spacetime by the non-geometric $R$-flux background.

To compute the coarse-graining quantitatively, we introduce oriented
area and volume uncertainty operators
\bea
\begin{aligned}
\A^{IJ} &= \mathrm{Im}\big([\widetilde x\,^I, \widetilde x\,^J]_\star\big) = -\ii\big({\widetilde x\,^I} \star {\widetilde x\,^J} - {\widetilde x\,^J} \star {\widetilde x\,^I}\big) \ , \\[4pt] \V^{IJK} &= \frac13 \, \mathrm{Re} \big( \widetilde
x\,^I\star [\widetilde x\,^J,\widetilde x\,^K]_{\star}+\widetilde
x\,^K\star [\widetilde x\,^I,\widetilde x\,^J]_{\star}+\widetilde
x\,^J\star [\widetilde x\,^K,\widetilde x\,^I]_{\star}\big) \ , 
\end{aligned}
\eea
where we introduced the shifted coordinates $\widetilde{x}\,^{I}:=
x^I-\langle x^I\rangle_\psi$ appropriate to uncertainties in measurements. These definitions mimick the vector
product and triple scalar product of vectors in the respective
coordinate directions of phase space, and their expectation values
measure the minimal area and volume uncertainties between
corresponding operators. A simple computation using
\eqref{eq:starcomm}, \eqref{eq:starJac} and the integration properties
above leads to the non-vanishing expectation values
\bea
\langle \A^{x^i,p_j}\rangle_\psi = \hbar\, \delta^i{}_j \ , \qquad
\langle \A^{ij}\rangle_\psi = \frac{\ell_s^4}{3\hbar} \, R^{ijk}\, \langle
p_k\rangle_\psi \qquad \mbox{and} \qquad \langle \V^{ijk} \rangle_\psi
= \frac12\, \ell_s^4\, R^{ijk} \ .
\eea
The first expectation value gives the usual Planck cells of canonical
quantum phase space with the Planck quantum of minimal area $\hbar$, while the
second expectation value is a new uncertainty measurement giving minimal spacetime
areas in the directions transverse to the motion of closed
strings (as originally conjectured by~\cite{stnag2}). The third expectation value is the most interesting: It
implies a quantized spacetime with a quantum of minimal volume 
$\frac12\, \ell_s^4\, R^{ijk}$.

This coarse-graining of spacetime has the following physical
interpretation from the perspective of non-geometric string theory,
which corroborates the argument that there can be no D0-branes in the locally
non-geometric background~\cite{Wecht:2007wu}. In $d=3$ dimensions, applying a triple T-duality $\Tcal_{(123)}$ maps the $R$-space to the three-torus with $H$-flux and a
D0-brane to a D3-brane wrapping the torus. However, this latter
configuration is not allowed as it suffers from the Freed-Witten
anomaly~\cite{FreedWitten}, i.e. it violates the Bianchi identity $\dd F=H$ for the gauge
flux $F$ on a D3-brane. Generally, Freed-Witten anomaly cancellation would
require that the degree three integer cohomology class of the NS--NS flux $H$ be
equal to the torsion characteristic class which measures the obstruction to a
spin$^c$ structure on the target space $M$. But if $M$ is a torus,
then its cohomology is torsion-free and it is a spin$^c$ manifold, so that non-vanishing $H$-flux is not
allowed. In the T-dual frame, this implies that placing a point-like
object in the $R$-flux background is \emph{not} allowed. This is yet
another manifestation of the local non-geometry of the $R$-flux
background, which we have reproduced here in a completely quantitative
way through our formalism of nonassociative quantum mechanics.

\subsection{Quantization of Nambu brackets}

As another application of the nonassociative phase space deformation
quantization, let us examine the problem of quantizing Nambu-Poisson
structures, which have notably appeared in recent years in effective
theories of M2-branes and M5-branes in M-theory, see
e.g.~\cite{Ho:2016hob} for a review. Nambu mechanics involves
multi-Hamiltonian dynamics with generalized Poisson brackets
$\{f,g,h\}$ of functions obeying the Leibniz rule and a ``fundamental
identity'' which is a higher generalization of the Jacobi identity for
Lie brackets~\cite{Nambu}. Nambu used this algebraic structure to
reformulate the Euler equations describing the dynamics of a rotating
rigid body in $\real^3$ in the absence of applied torques, in the hope
of obtaining new generalized integrals of motion. Recall that in a rotating reference frame parallel to the principal axes of inertia of the rigid body, they read in general form as
\bea
\dot{\vec L}+ \vec\omega\times\vec L=\vec0 \ ,
\eea
where $\vec L=\mbf I\, \vec\omega$ is the angular momentum, with $\mbf I$ the matrix of moments of inertia and $\vec\omega$ the angular velocity about the principal axes. By setting $T:=\frac12\, \vec L\cdot\vec\omega$ we can write the components of this equation in the principal reference frame, wherein $\mbf I$ is constant, as the bi-Hamiltonian equations
\bea
\dot L_i = \{L_i,\vec L{}^{\,2},T\}
\eea
for the Nambu bracket $\{f,g,h\} = \varepsilon^{ijk}\, \partial_{L_i}f\cdot \partial_{L_j}g\cdot \partial_{L_k}h$ on $\real^3$.

In the case of the $R$-flux background, the pertinent classical tribracket is given by
\bea\label{eq:Rfluxtribracket}
\{f,g,h\}_R = - \frac{\ell_s^4}{\hbar^2}\, R^{ijk} \, \partial_if\cdot\partial_jg\cdot\partial_kh \ .
\eea
Our nonassociative phase space star product quantizes these tribrackets~\cite{Mylonas2012,Mylonas2013}, in the sense of the non-vanishing Jacobiators \eqref{eq:starJac}; this extends the quantization provided by the configuration space triproducts from \eqref{eq:triprodbracket}. In particular, in the semi-classical limit one has
$[f,g,h]_\star = -\hbar^2 \, \{f,g,h\}_R + \cdots$.
The issue of quantizing Nambu brackets is a longstanding problem, see e.g.~\cite{SheikhJabbari:2004ik,SheikhJabbari:2005mf,DeBellis:2010pf,DeBellis:2010sy} for some approaches. Our phase space approach may assist by exploiting the various properties infered by the origin of the tribracket from the star product. For example, one of the formality conditions implies the pentagon identity~\cite{Mylonas2012}
\bea
[f\star g,h,k]_\star-[f,g\star h,k]_\star + [f,g,h\star k]_\star = f\star[g,h,k]_\star+[f,g,h]_\star\star k
\eea
for the Jacobiator, analogously to \eqref{eq:pentagon}.
This can be regarded as a `quantum Leibniz rule', in the sense that at semiclassical order in the $R$-flux it coincides with the Leibniz rule for the classical tribracket \eqref{eq:Rfluxtribracket}. At present it is not clear what should serve as the quantum version of the classical fundamental identity, nor if one is really necessary, see e.g.~\cite{DeBellis:2010pf} for a discussion of this point. The viability of our phase space model for the quantization of Nambu-Poisson structures is still to be thoroughly investigated.

%%%%%%%%%%%%%%%%%%%%%%%%%%%%%%%%%%%%%%%%%%%%%%%%%%%%%%%
%%%%%%%%%%%%%%%%%%%%%%%%%%%%%%%%%%%%%%%%%%%%%%%%%%%%%%%

\section{Further developments}
\label{sec:further}

We conclude in this final section by briefly describing some extensions of the story
described in this paper so far, all of which have important open problems
which should be subject to future investigations.

\subsection{Nonassociative gravity}

One important omission from our considerations thus far has been
addressing Question~(Q3) from Section~\ref{sec:intro}. A
noncommutative theory of gravity on Moyal-Weyl spacetimes was
constructed in~\cite{Aschieri:2005yw}, and is discussed in the lectures by L.~Castellani at this School. As a fundamental length is incompatible with diffeomorphism symmetries, general relativity on noncommutative spacetime requires a Drinfeld twist via a two-cocycle of the Hopf enveloping algebra of the Lie algebra of vector fields (see e.g.~\cite{Szabo:2006wx}
for a review). This defines a twisted tensor calculus, and leads to deformed Einstein equations~\cite{Aschieri:2005yw}. Two major problems with this approach are that the Einstein equations are not generically real and so have questionable physical (and geometric) significance, and that twisted diffeomorphisms do not appear to be symmetries of string theory~\cite{AlvarezGaume:2006bn}.

An analogous treatment can be followed to formulate a nonassociative theory of gravity, by defining a quasi-Hopf algebra of twisted diffeomorphisms with a two-cochain twist whose coboundary is the three-cocycle 
that controls nonassociativity, i.e. the associator $\varphi$, see \eqref{eq:associator}. Such an approach was originally suggested by~\cite{Mylonas2013}, and the pentagon relations \eqref{eq:pentagon} were used to consistently build nonassociative field theories in~\cite{Mylonas:2014kua,Barnes:2016cjm}. The cochain twisting method was extended by~\cite{Barnes:2014ksa,Barnes:2015uxa} to develop a rigorous and very general theory of nonassociative differential geometry, extending and generalizing the considerations of~\cite{AschieriSchenkel} in the associative case, which in particular leads to a 
vielbein or first order formalism for nonassociative gravity. The metric aspects were considered in~\cite{Blumenhagen:2016vpb,Aschieri:2017sug}, and currently nonassociative Riemannian geometry is under further construction.

The problems with noncommutative gravity alluded to above are of course still present in this theory (together with many additional issues). But recalling the development of our perspective, we need to explore how gravity on phase space induces gravity on configuration space. This problem was addressed by~\cite{Aschieri:2015roa}, where it is shown how the nonassociative geometry of phase space can lead to deformations of configuration space geometry. The idea is to remove momentum dependence up to an $O(d,d)$-transformation, which is the structure group symmetry determined by the natural phase space metric $\gamma =\dd x^i\otimes \dd p_i+\dd p_i\otimes \dd x^i$, analogously to double field theory. This is achieved by choosing a polarization, i.e. a maximally isotropic splitting $T(T^*M) \simeq L\oplus L^*$ of the tangent bundle of phase space with respect to the metric $\gamma$, and considering foliated tensor fields $T$:
\bea
\imath_ZT=0=\pounds_ZT \ ,
\eea
for every section $Z$ of the rank $d$ vector bundle $L^*\to T^*M$; if the distribution $L$ is integrable, then by Frobenius' theorem it can be (locally) identified with the tangent bundle of some $d$-dimensional submanifold of $T^*M$. For example, the foliation of phase space by leaves of constant momentum is determined by taking $L=TM$ and $Z_i=\frac\partial{\partial p_i}$ for $i=1,\dots,d$, which formalizes the way in which we extracted the spacetime triproducts of fields occuring in closed string scattering amplitudes from the nonassociative phase space star product in \eqref{eq:tristarproducts}.

The metric formulation of nonassociative gravity on phase space admits a Ricci tensor and a unique metric-compatible torsion-free connection, and in this way it
yields a non-trivial deformation of the Ricci tensor of spacetime by locally non-geometric fluxes, which is given by~\cite{Aschieri:2017sug}
\bea
\begin{aligned}
{\rm Ric}^\circ_{ i j} &= {\rm Ric}_{ i j} +
\frac{\ell_s^4}{12} \, R^{ a b c}\,\Big(
\partial_ k\big(\partial_ a  g^{ k l}\, (\partial_ b
  g_{ l m})\, \partial_ c \Gamma_{ i j}^{m} \big) 
- \partial_ j\big(\partial_ a  g^{ k l}\, (\partial_ b
  g_{ l m})\, \partial_ c \Gamma_{ i k}^{  m}\big)
\\ & \qquad \qquad \qquad \qquad +\,  \partial_ c   g_{ m n}\big( \partial_ a
(  g^{ l m}\, \Gamma_{ l j}^{ k})\, \partial_ b
\Gamma_{ i k}^{ n} - \partial_ a
(  g^{ l m}\, \Gamma_{ l k}^{ k})\,\partial_ b
\Gamma_{ i j}^{ n}
\\ & \qquad\qquad \qquad \qquad +\, (\Gamma_{ i k}^{ l}\, \partial_ a
  g^{ k m} - \partial_ a \Gamma_{ i k}^{ l} \,  g^{ k m})\,           
\partial_ b \Gamma_{ l j}^{n} - (\Gamma_{ i j}^{ l}\, \partial_ a
  g^{ k m} -\partial_a \Gamma_{ i j}^{ l}\,  g^{ km})\,              
\partial_b \Gamma_{l k}^{n} \big) \Big) \ ,
\end{aligned}
\label{eq:Ricciintro}\eea
where ${\rm Ric}_{ i j}$ is the usual Ricci tensor of the classical
Levi-Civita connection $\Gamma_{ i j}^{k}$ of a metric
tensor $g_{ ij}$ on spacetime. This expression is valid to linear
order in the $R$-flux, which we recall is the order at which the corresponding
conformal field theory calculations are
reliable; this is also consistent with the second order $R$-flux
corrections to the closed string equations \eqref{eq:SUGRAeqs} after
T-duality, as the corrections \eqref{eq:Ricciintro} are indeed of
second order $\ell_s^4\, \hbar$ in the double expansion in the
parameters of the $R$-flux model. Notably, it is \emph{real}, and thus represents the first non-trivial
starting point for understanding how to define
a nonassociative theory of gravity describing the low-energy effective
dynamics of closed strings in non-geometric backgrounds. An action
principle for nonassociative gravity is currently unknown, as some of
the classical constructions of Riemannian geometry have yet to be generalized to the nonassociative setting.

The precise relation of this gravity theory with string theory and double field theory remains mysterious. Part of the issue is the role of twisted diffeomorphisms mentioned above. At any order in the string length scale $\ell_s^2$, the closed string effective action should be invariant under classical diffeomorphisms, while the effective action of double field theory should be invariant under generalized diffeomorphisms. Whereas the twisted diffeomorphism symmetries of phase space remain elusive, upon polarization to configuration space they may compare naturally with the expectations from string theory and double field theory. This was partially analysed by~\cite{Aschieri:2015roa}, but it is currently an open problem to understand precisely and systematically the meaning of the symmetries of nonassociative gravity on spacetime. In particular, one can ask if the effective theory retains the $O(d,d)$-symmetry of double field theory, in analogy to the case of open strings. Recall~\cite{Douglas:2001ba,Szabo:2001kg} that noncommutative Yang-Mills theory on a $d$-dimensional torus is invariant under Morita equivalence, which is the structure inherited from T-duality in the decoupling limit. In this case the group $O(d,d)$ acts on the bivector $\theta$, the open string metric $G$, and the Yang-Mills coupling constant $g_{\rm YM}$ as
\bea
\begin{aligned}
\theta &\longmapsto (A\,\theta+B)\, (C\,\theta+D)^{-1} \ , \\[4pt]
G &\longmapsto (C\,\theta+D)\, G\, (C\,\theta+D)^\top \ , \\[4pt]
g_{\rm YM} &\longmapsto g_{\rm YM}\, \big|\det(C\,\theta+D)\big|^{1/4} \ ,
\end{aligned}
\eea
where $A,B,C,D$ are $d\times d$ matrices which parameterize an element of $O(d,d)$; this transformation is defined only on the subset of bivectors $\theta$ for which $C\,\theta+D$ is nondegenerate, which is dense for every element in the discrete subgroup $SO(d,d;\zed)$ of $O(d,d)$. It would be interesting to see if a similar set of T-duality transformation rules are global symmetries of nonassociative gravity.

\subsection{Higher structures in non-geometric M-theory}

It is natural to ask what becomes of the higher quantum geometry when we lift IIA string theory to M-theory, which is generally defined on a circle bundle
\bea
\xymatrix{
S^1 \ \ar@{^{(}->}[r] & \ \widetilde{M} \ar[d] \\
 & \ M
}
\eea
over the string target space $M$, with the string coupling $g_s$ realized geometrically as the radius $\lambda$ of the $S^1$ fibres. This has been discussed for lifts of the string theory $R$-flux model in three dimensions by~\cite{Gunaydin:2016axc}, in the context of the $SL(5)$ exceptional field theory~\cite{Blair:2014zba} which lifts the $O(3,3)$ double field theory. Taking the base space $M$ to be a three-dimensional twisted torus (the T-dual of the three-torus with $H$-flux), using the T-duality chain \eqref{eq:Tdualitychain} we generate the string theory $R$-flux via a double T-duality transformation 
\bea
f^i{}_{jk} \xrightarrow{ \ \Tcal_{(jk)} \ } \ R^{ijk} \ .
\eea
Let us lift this to M-theory on the trivial circle bundle $\widetilde{M} =M\times S^1$, with coordinates $x^\mu=(x^i,x^4)$ where $x^4$ is the local coordinate on $S^1$. Closed strings lift to closed M2-branes, and T-duality becomes U-duality which sends membrane wrapping modes $ w^{ij}$ to momentum modes $p_i$. 
The Kalb-Ramond two-form field $B$ lifts to the three-form $C$-field of M-theory, and U-duality takes it to
\bea
C_{\mu\nu\rho}
  \xrightarrow{ \ \CU_{(\mu\nu\rho)} \ }\Omega^{\mu\nu\rho} \ ,
\eea
where the trivector $\Omega^{\mu\nu\rho}$ defines the M-theory $R$-flux via its wrapping derivatives
\bea
R^{\mu,\nu\rho\alpha\beta}=\partial^{\mu[\nu}\Omega^{\rho\alpha\beta]} \ ,
\eea
which is the lift of the relation \eqref{eq:Rfluxtheta}. The M-theory $R$-flux is a mixed symmetry tensor: It transforms as a vector in its first index and is antisymmetric in its last four indices. The particular choice $R^{4,\mu\nu\alpha\beta} = R\,
    \varepsilon^{\mu\nu\alpha\beta}$ breaks the $SL(5)$ symmetry to $SO(4)$.
    
The M2-brane phase space has a peculiar structure: The fact that there are no D0-branes on $M$ lifts to the statement that there are no momentum modes along the M-theory direction, i.e. $p_4=0$. In~\cite{Gunaydin:2016axc} it is conjectured that the covariant form of this constraint is given by
\bea
R^{\mu,\nu\rho\alpha\beta}\, p_\mu=0
\eea
and that the resulting seven-dimensional phase space has bracket structure given by
\bea\label{eq:M2octonion}
\begin{aligned}
{} [x^i,x^j] &=\frac{\ii\ell_s^4}{3\hbar} \,
                        R^{4,ijk4}\, p_k  \qquad \mbox{and} \qquad
                        [x^4,x^i] =\frac{\ii \lambda\, \ell_s^4}{3\hbar} \, R^{4,1234}\, p^i  \ , \\[4pt]
{} [x^i,p_j] &=  \ii\hbar\, \delta^i{}_j\,x^4+\ii\hbar\, \lambda\,
                        \varepsilon^i{}_{jk}\, x^k \qquad \mbox{and}
                        \qquad [x^4,p_i] = \ii\hbar\, \lambda^2\,x_i \ , \\[4pt]
{} [p_i,p_j] &= -\ii\hbar\,\lambda\, \varepsilon_{ijk}\, p^k \ , 
\end{aligned}
\eea
with the Jacobiators
\bea
\begin{aligned}
{} [x^i,x^j,x^k] &= \frac{\ii\ell_s^4}{3\hbar}\,
                            R^{4,ijk4} \, x^4 \qquad \mbox{and}  \qquad
{} [x^i,x^j,x^4] = -\frac{\ii\lambda^2\, \ell_s^4}{3\hbar} \, R^{4,ijk4} \, x_k \ , \\[4pt]
 {} [p_i,x^j,x^k] &= \frac{\ii\lambda\,\ell_s^4}{3\hbar} \, R^{4,1234}\, \big(\delta_i{}^j\, p^k-\delta_i{}^k\, p^j \big) \qquad \mbox{and} \qquad
{} [p^i,x^j,x^4]  =  \frac{\ii\lambda^2\, \ell_s^4}{3\hbar} \, R^{4,ijk4}\, p_k \ , \\[4pt]
 {} [p_i,p_j,x^k]  &= -\ii\hbar \, \lambda^2\,
                            \varepsilon_{ij}{}^{k}\, x^4-\ii\hbar\, \lambda\, \big(\delta_j{}^k\, x_i-\delta_i{}^k\, x_j \big) \qquad \mbox{and} \qquad
{} [p_i,p_j,x^4] = \ii\hbar\, \lambda^3\, \varepsilon_{ijk}\, x^k \ , \\[4pt]
 {} [p_i,p_j,p_k] &= 0 \ .
\end{aligned}
\eea
In the contraction limit $\lambda=0$, which is precisely the limit of weak string coupling $g_s\to 0$ sending M-theory to IIA string theory, these brackets reduce to those of the closed string $R$-flux algebra \eqref{eq:Rfluxalgebra}; in this case the M-theory direction $x^4$ becomes a central element of the contracted algebra, so we may set it to $x^4=1$.

These brackets originate from the nonassociative alternative algebra of octonions $\mathbb{O}$ by a suitable rescaling of the seven imaginary unit octonions~\cite{Gunaydin:2016axc}. Deformation quantization of this quasi-Poisson structure was carried out in~\cite{Kupriyanov:2017oob}. It proceeds via a choice of a $G_2$-structure, i.e. a cross product on the real inner product space $\real^7$ given by structure constants of $\mathbb{O}$ in a suitable oriented basis, which is preserved by rotations of $\real^7$ in the subgroup $G_2\subset SO(7)$. Using alternativity one can define octonion
  exponentials, and the corresponding nonassociative Baker-Campbell-Hausdorf formula is captured by a 2-group addition law on the seven-dimensional Fourier space, which gives the deformation quantization of the M2-brane phase space in terms of an explicit, albeit complicated, nonassociative phase space star product $f\star_\lambda g$. This quantization has a variety of interesting features which are described in~\cite{Kupriyanov:2017oob}. For example, the corresponding configuration space triproducts $\tri_\lambda$ defined by $\star_\lambda$ analogously to \eqref{eq:tristarproducts} quantize the 3-Lie algebra $A_4$:
  \bea
  [x^\mu,x^\nu,x^\alpha]_{\vartriangle_1} = \ell_s^4\,R
\,\varepsilon^{\mu\nu\alpha\beta}\, x^\beta \ ,
  \eea 
which is known from studies of multiple membranes in M-theory, see e.g.~\cite{Bagger2012} for a review, and whose quantization has been previously largely unknown, see e.g.~\cite{DeBellis:2010pf} for an earlier analysis of this problem. Moreover, the noncommutative M2-brane momentum space is familiar from the noncommutative spacetimes arising in three-dimensional quantum gravity, see e.g.~\cite{Freidel2005}. Using $x^4\star_\lambda f=x^4\,f+O(\lambda)$ one finds that the M2-brane star product reduces non-trivially to the closed
  string star product \eqref{eq:NAstarproduct} in the weak string coupling limit:
\bea
\lim_{\lambda\to0} \, (f\star_\lambda g)(x^i,x^4) = (f\star g)(x^i) \ .
\eea

In~\cite{Kupriyanov:2017oob} it was also suggested how to extend the nonassociative M-theory $R$-flux algebra to the full unconstrained eight-dimensional phase space by a choice of a $Spin(7)$-structure, i.e. a triple cross product on the real inner product space $\real^8$, which is parameterized by a four-form $\phi$ invariant under rotations of $\real^8$ in the subgroup $Spin(7)\subset SO(8)$. This extends the representation of the cross product to all real and imaginary octonions, and it defines the eight-dimensional ``covariant'' M2-brane phase space 3-algebra with the $SO(4){\times}SO(4)$-symmetric 3-brackets given by
\bea\label{eq:M23algebra}
\begin{aligned}
{} [x^i,x^j,x^k]_\phi  &= -\frac{\ell_s^4}{2} \,
R^{4,ijk4}\, x^4  \qquad \mbox{and} \qquad {} [x^i,x^j,x^4]_\phi =
\frac{\lambda^2\,\ell_s^4}{2}\, R^{4,ijk4}\, x_k  \ , 
\\[4pt] {} [p^i,x^j,x^k]_\phi &=
-\frac{\lambda^2\,\ell_s^4}{2}\, R^{4,ijk4}\,
p_4-\frac{\lambda\,\ell_s^4}{2}\, R^{4,ijkl}\, p_l
\ , \\[4pt]
{} [p^i,x^j,x^4]_\phi &=
 -\frac{\lambda^2\,\ell_s^4}{2} \, R^{4,1234}\,
\delta^{ij} \, p_4-\frac{\lambda^2\,\ell_s^4}{2} \,
R^{4,ijk4}\, p_k \ , \\[4pt]
 {} [p_i,p_j,x^k]_\phi &= \frac{\lambda^2}2 \,
\varepsilon_{ij}{}^k\, x^4+ \frac{\hbar^2\,\lambda}2 \,\big(\delta_j{}^k\,
x_i-\delta_i{}^k\, x_j\big)   \ , \\[4pt]
 {} [p_i,p_j,x^4]_\phi &= - \frac{\hbar^2\, \lambda^3}2 \,
\varepsilon_{ijk}\, x^k \qquad \mbox{and} \qquad 
{} [p_i,p_j,p_k]_\phi \ = \ -2\hbar^2\,\lambda \,
\varepsilon_{ijk}\, p_4   \ , \\[4pt]
{} [p_4,x^i,x^j]_\phi &= \frac{\lambda\, \ell_s^4}{2}  \,
R^{4,ijk4}\, p_k \qquad \mbox{and} \qquad {} [p_4,x^i,x^4]_\phi   =
  - \frac{\lambda^2\, \ell_s^4}{2} \, R^{4,1234}\, p^i
\ , \\[4pt]
 {} [p_4,p_i,x^j]_\phi &= - \frac{\hbar^2\,\lambda}2 \, \delta_i{}^j\,
x^4- \frac{\hbar^2\, \lambda^2}2  \, \varepsilon_i{}^{jk}\, x_k  \ ,
\\[4pt]
{}  [p_4,p_i,x^4]_\phi &= - \frac{\hbar^2\,\lambda^3}2  \, x_i
                          \qquad \mbox{and} \qquad
{} [p_4,p_i,p_j]_\phi =
- \frac{\hbar^2\, \lambda^2}2  \, \varepsilon_{ijk}\, p^k \ .
\end{aligned}
\eea
For any constraint $G=0$ on the eight-dimensional phase space, these 3-brackets induce a 2-bracket by defining $[f,g]_G:= [f,g,G]_\phi$. In particular, taking $G =\frac2{\lambda\,\hbar^2}\, p_4$ (or $G =
  R^{\mu,\nu\rho\alpha\beta} \, p_\mu$) gauge fixes the 3-brackets \eqref{eq:M23algebra} to the brackets \eqref{eq:M2octonion} together with their Jacobiators. The quantization of this 3-algebra is currently an open problem; see~\cite{Kupriyanov:2017oob} for some preliminary steps towards the construction of a suitable ternary product for deformation quantization which naturally incorporates both the star product and the triproduct.

The magnetic monopole system of Section~\ref{sec:monopole} with constant uniform magnetic charge can be embedded in string theory as D0-branes bound to a uniform distribution of D6-branes. The M-theory lift of this configuration is identified by~\cite{Lust:2017bgx} as a non-geometric variant of the Kaluza-Klein monopole solution of M-theory, whose phase space brackets map to \eqref{eq:M2octonion} under magnetic duality and reduce to the magnetic monopole algebra \eqref{eq:monopolealgebra} in the contraction limit $\lambda=0$.

A suitable check of these purported claims should come from the next rung $n=3$ on the AKSZ geometric ladder of Section~\ref{sec:geometry}, which has been investigated partially in~\cite{Kokenyesi:2018ynq}; the symplectic Lie 3-algebroid structure in this case is called a \emph{Lie algebroid up to homotopy}, and the corresponding AKSZ sigma-model describes closed M2-branes viewed as boundaries of open threebranes. The standard Lie algebroid up to homotopy is the anticipated generalized tangent bundle $T\widetilde{M}\oplus\bigwedge^2T^*\widetilde{M}$ of exceptional field theory~\cite{dftrev2}, and the bracket structure in this case is the expected higher 2-Courant bracket. However, beyond the simplest case of the four-form flux $G=\dd C$ of the M-theory $C$-field, it is not clear how to twist these higher algebroid structures by geometric and non-geometric fluxes; see~\cite{Kokenyesi:2018ynq} for a discussion of this point. This may be related to the fact that, unlike T-duality which maps closed strings to closed strings, U-duality maps M2-branes to M-waves (the lifts of D0-branes), and more generally to M5-branes in higher dimensions.

Locally non-geometric fluxes in M-theory on higher dimensional spacetimes have been discussed in~\cite{Lust:2017bwq}. In each dimensionality one encounters the same qualitative structure, missing momentum modes in the M2-brane phase space, but now with a host of $R$-type fluxes with varying tensorial structures. The corresponding bracket structures are not currently understood, and may also involve higher algebras.

\subsection{$L_\infty$-algebras}

As mentioned in Section~\ref{sec:quantization}, the Kontsevich formality maps are quasi-isomorphisms between particular examples of $L_\infty$-algebras, and it is natural to wonder if such higher algebraic structures govern the infinitesimal symmetries of the higher quantum geometries we have discussed. In~\cite{Mylonas2012} it was observed that the nonassociative phase space deformation \eqref{eq:Rfluxalgebra} underlying the closed string $R$-flux background can be realized and understood as a 2-term $L_\infty$-algebra, which is similar to the characterization~\cite{Roytenberg:1998vn} of the bracket structure underlying Courant algebroids from Section~\ref{sec:geometry}. This result was extended, and put into a more general and systematic framework by~\cite{Hohm:2017cey}, showing in particular how the M-theory $R$-flux algebra \eqref{eq:M2octonion} is realized in the same way. It would be interesting to similarly understand the covariant M2-brane 3-algebra \eqref{eq:M23algebra} in this way, which may also be the appropriate framework for the higher bracket structures underlying the phase spaces of non-geometric M-theory in dimensions $d>4$ where they are governed by the $E_d$ exceptional field theory~\cite{Lust:2017bwq}. 

These occurences nicely match the original appearence of $L_\infty$-algebras in physics as higher gauge symmetry algebras of closed string field theory in~\cite{Zwiebach1992}, and more recently as the higher symmetries underlying double field theory~\cite{Deser:2014mxa,Deser:2016qkw,Hohm:2017pnh} and two-dimensional conformal field theory~\cite{Blumenhagen:2017ogh}. It was shown by~\cite{Blumenhagen:2018kwq} that the symmetries and dynamics of open string nonassociative gauge theories on curved D-brane worldvolumes are also governed by an underlying $L_\infty$-algebra. Extrapolating this feature to non-geometric backgrounds could lead to a similar characterization of the diffeomorphism symmetries in nonassociative gravity on phase space, and its fate under polarization to spacetime which may clarify the connections with closed string theory and double field theory discussed above.

%%%%%%%%%%%%%%%%%%%%%%%%%%%%%%%%%%%%%%%%%%%%%%%%%%%%%%%
%%%%%%%%%%%%%%%%%%%%%%%%%%%%%%%%%%%%%%%%%%%%%%%%%%%%%%%

\acknowledgments{
We would like to thank the organisors of the School for the invitation
to deliver these lectures, and all the participants for their
persistent questions and comments which have helped shape the
structure and content of these lecture notes. Several points of view
discussed in this article have emerged from numerous discussions over
the years with various collaborators and colleagues; we would like to
thank in particular Paolo Aschieri, Ralph Blumenhagen, Chris Hull,
Dieter L\"ust, Vladislav Kupriyanov, Emanuel Malek, Alexander Schenkel and Peter Schupp.
This work was supported in part by the Action MP1405 QSPACE 
from the European Cooperation in Science and Technology
(COST) and by the Consolidated Grant ST/P000363/1
from the UK Science and Technology Facilities Council (STFC). 
}

%%%%%%%%%%%%%%%%%%%%%%%%%%%%%%%%%%%%%%%%%%%%%%%%%%%%%%%
%%%%%%%%%%%%%%%%%%%%%%%%%%%%%%%%%%%%%%%%%%%%%%%%%%%%%%%

\end{document}